\documentclass[showpacs,preprintnumbers,aps]{revtex4}
\usepackage{graphicx}
\usepackage{dcolumn}
\usepackage{bm}

\begin{document}

\title{Equations of motion approach to the spin-$1/2$ Ising model on the Bethe lattice}
\author{Ferdinando Mancini}
\email{mancini@sa.infn.it}
\author{Adele Naddeo}
\email{naddeo@sa.infn.it}

\affiliation{Dipartimento di Fisica "E.
R. Caianiello" - Unit\'{a} CNISM di Salerno, Universit\'{a} degli
Studi di Salerno, 84081 Baronissi (SA), Italy}

\date{\today}

\begin{abstract}
We exactly solve the ferromagnetic spin-$1/2$ Ising model on the
Bethe lattice in the presence of an external magnetic field by
means of the equations of motion method within the Green's
function formalism. In particular, such an approach is applied to
an isomorphic model of localized Fermi particles interacting via
an intersite Coulomb interaction. A complete set of eigenoperators
is found together with the corresponding eigenvalues. The Green's
functions and the correlation functions are written in terms of a
finite set of parameters to be self-consistently determined. A
procedure is developed, that allows us to exactly fix the unknown
parameters in the case of a Bethe lattice with any coordination
number $z$. Non-local correlation functions up to four points are
also provided together with a study of the relevant thermodynamic
quantities.
\end{abstract}

 \maketitle



\section{Introduction}

Recently, it has been shown \cite{professore1} that a system built
up of $q$ species of Fermi particles, localized on the sites of a
Bravais lattice and subjected to finite-range interactions, is
exactly solvable in any dimension. Exactly solvable means that it
is always possible to find a complete set of eigenvalues and
eigenoperators of the Hamiltonian, which close the hierarchy of
the equations of motion. In such a way, exact expressions for the
relevant Green's functions and correlation functions can be
derived. These expressions are just formal because they depend on
a finite set of parameters to be self-consistently determined. In
Refs. \cite {professore2,professore3,professore4} it has been
shown how it is possible to fix such parameters exactly by means
of algebra constraints in the case of one dimension and $q=1$,
$q=2$ and $q=3$, respectively. In this way, complete and exact
solutions of these systems have been obtained.

A system of $q$ species of Fermi particles has been shown
\cite{professore1} to be isomorphic to a spin-$\frac{q}{2}$
Ising-like model in the presence of an external magnetic field, so
opening a different route to the study of spin systems, which can
be very difficult in two and three dimensions when attached by the
transfer matrix method. Furthermore, this approach can shed new
light on how to get an exact solution for these systems in higher
dimensions in the presence of an external magnetic field as it is
always possible \cite{professore1} to find an exact expression for
the corresponding Green's functions and correlation functions. The
exact knowledge of the eigenenergies of the system can give
information on the energy scales ruling the physical behavior and
the response of the system and can find an application as unbiased
check for the approximate studies present in the literature.
Within our approach the problem is that correlation and Green's
functions depend on a finite set of unknown parameters to be
self-consistently determined. A complete exact solution of the
system is obtained only when such parameters are known. These
parameters cannot be determined by means of the dynamics and are
fixed by choosing the representation where the field operators are
realized. In particular [cfr. Section 2.4 in Ref.
\cite{professore6}], they can be fixed by appropriate
self-consistent equations which are the manifestation of
symmetries of the model, algebraic properties of the field
operators, boundary conditions (i.e. properties of the underlying
lattice, phase of the system according to the values of the
external thermodynamical parameters). It has been shown how to fix
exactly such parameters by means of algebra constraints in the one
dimensional (1D) case \cite{professore2,professore3,professore4}
and we are now working on the possibility to introduce new algebra
constraints and topological relations in order to fix the
self-consistent parameters in higher dimensions. The first step to
realize such an ambitious program appears to be the application of
our technique to a Bethe lattice of any coordination number $z$.
In this article we apply our formulation, the Composite Operator
Method (COM) \cite{professore5,professore6}, to the study of a
system of $q=1$ species of Fermi particles interacting with an
intersite Coulomb interaction on the Bethe lattice with
coordination number $z$. Such a model is shown to be isomorphic to
the spin-$\frac{1}{2}$ Ising model on the Bethe lattice in the
presence of an external magnetic field.

Bethe \cite{domb} and Bethe-like lattices \cite{essam} have been
widely studied in solid state and statistical physics because they
represent the underlying lattices on which many systems can be
exactly solved \cite
{runnels,eggarter,mullerh,hu1,chack,izmail1,efetov,yokoi,gujrati,kopec,stanley,mirlin,monroe}%
. Following the line of reasoning that refers to the mean field
theory of magnetism, the Bethe-Peierls \cite{bethep} approximation
was developed in 1935/36 in order to describe crystalline alloys
or Ising models. It takes exactly into account the interaction of
a given spin with its nearest neighbors but introduces a mean
field in order to express the interactions between such neighbors
and all the other spins in the lattice. Later, it was pointed out
\cite{kurata} that such an approximation becomes exact on the
Bethe lattice. In particular, it was shown that the partition
function of the ferromagnetic spin-$\frac{1}{2}$ Ising model on
the Bethe lattice with any coordination number $z$ is equivalent
to that in the Bethe approximation \cite{kurata}. Then, the
equivalence of the exact solution of the Bethe lattice
spin-$\frac{1}{2}$ Ising model to the Bethe-Peierls approximation
was shown also in the antiferromagnetic case \cite{katsura}.
Summarizing, there are two special properties that make Bethe
lattices particularly suited for theoretical investigations: the
self-similar structure which may lead to recursive solutions and
the absence of closed loops which restricts interference effects
of quantum-mechanical particles in the case of nearest-neighbor
coupling. Furthermore, Bethe and Bethe-like lattices have
attracted a lot of interest because they usually reflect essential
features of systems even when conventional mean-field theories
fail \cite{gujrati}. The reason is that such lattices are capable
to take into account correlations which are usually lost in
conventional mean-field calculations. The spin-$\frac{1}{2}$ Ising
model on the Bethe lattice can be exactly solved by means of the
transfer matrix technique \cite{baxter} which reduces the solution
to an eigenvalue problem of the second order and all the relevant
thermodynamic quantities such as the magnetization and the
susceptibility, and so on, can be calculated by means of recursion
relations, thanks to the nested structure of the underlying
lattice. The same technique has been recently employed in the
exact calculation of the spin-spin correlation functions
$\left\langle S\left( 0\right) S\left( n\right) \right\rangle $
for any temperatures $T$ and external field $h$
\cite{izmail2,goltsev}. Exact expressions for the free energy and
the magnetization of a spin-$\frac{1}{2}$ Ising model on a
two-layer Bethe lattice\ in the presence of magnetic fields
different in the two layers have been also obtained together with
a study of the whole phase diagram by means of an iteration
technique \cite{izmail3}. Also, it should be mentioned a large
activity in the framework of the athermal random-field Ising model
(RFIM), where analytical results have been obtained on Bethe
lattices \cite{rfi1,rfi2}.

In this paper we exactly solve the ferromagnetic
spin-$\frac{1}{2}$ Ising model on the Bethe lattice with any
coordination number $z$ in the presence of an external magnetic
field within the COM approach \cite {professore5,professore6}. All
the Green's functions and correlation functions are obtained
together with the behavior of the relevant thermodynamic
properties. Two-point $\left\langle S\left( 0\right) S\left(
j\right) \right\rangle $ and three-point $\left\langle S\left(
0\right) S\left( j\right) S\left( k\right) \right\rangle $
spin-spin correlation functions are also provided together with
non local correlation functions of higher order. The manuscript is
organized as follows. In Section 2, we give the general
Hamiltonian of the spin-$\frac{1}{2}$ Ising model on the Bethe
lattice and the mapping onto a model of Fermi particles with
intersite Coulomb interactions. In Section 3, we present the
general solution in terms of eigenvalues and eigenvectors. In
Section 4, we show how to close the system of self-consistent
equations and find the unknown parameters in order to compute the
correlation and the Green's functions. In Section 5, we compute
the local correlation functions and in Section 6, the non-local
ones. In Section 7, we study all the relevant thermodynamic
quantities, such as magnetization, susceptibility, internal
energy, specific heat and entropy as functions of the temperature
and the external magnetic field, specializing the general formulas
to the case of a Bethe lattice with coordination number $z=3$ and
$z=4$. Finally, some concluding remarks and outlooks of our work
are given. Some technical Appendices follow.

As a final remark, we would like to stress that the motivation of
this work is to show that the formalism of Green's functions and
equations of motion is a convenient  technique to study spin
systems. Most of the techniques used in the literature for the
study of these systems are based on the transfer matrix method.
This latter formalism is a very powerful technique and has been
largely applied with success to a huge number of models. After the
brilliant solution by Onsager \cite {onsager} for the
two-dimensional spin-$\frac{1}{2}$ Ising model in zero field, many
other two-dimensional (2D) models, such as the dimer problem,
six-vertex,  eight-vertex  (see Baxter's book),  have been solved
by making use of the transfer matrix method [see Baxter's book
\cite {baxter} for a comprehensive list of references]. However,
it should be noticed that this method is very transparent and
convenient for the case of one dimension, but becomes complicate
for higher dimensions. In spite of the tremendous work done [among
the most recent results, the derivation of the order parameter of
the chiral Potts model by Baxter \cite {baxter1} has to be
mentioned], many problems remain unsolved. The exact partition
function in a finite magnetic field is still unknown. No exact
results have been obtained for the three-dimensional model. By
using the equation of motion formalism, we have constructed a
general method to study Ising spin systems \cite {professore1}.
Such a method is general, in the sense that it has been formulated
for any dimension of the system. We can exactly calculate a
complete set of eigenoperators and eigenvalues of the Hamiltonian,
and consequently to derive analytical expressions for the
correlation functions. In order that this scheme of calculation
could be used in practice, it is necessary to calculate a set of
unknown parameters. The number of unknown parameters depends not
only on the dimensions of the system, but also on the dimension of
the spin; for a Ising spin-$q/2$  system on a lattice of
coordination number $z$, the number of unknown parameters is $2qz$
[cfr. Ref. \cite {professore1}]. Our previous studies \cite
{professore1,professore2,professore3,professore4} show that it is
possible to find the necessary self-consistent equations by using
not only properties of the lattice, but also symmetry and
algebraic properties of the field operators. In the last two years
we have been  performing a systematic study of this last point. We
started by considering the simplest problem of spin-$1/2$ on a
linear chain \cite {professore2}. Then, we considered the case of
spin-1 \cite {professore3} and spin-$3/2$ \cite {professore4},
always for 1D systems. The extension to spin higher than $1/2$ is
not immediate, but requires the introduction of higher composite
fields (projection operators). After this study of 1D systems, we
decided to consider more complicate lattices, by considering the
Bethe lattice. This lattice has the same topology of 1D because
the absence of closed loops, but the analysis requires a
dependence on the coordination number. For the Bethe lattice we
have shown that the problem can be completely solved; we have
shown that all the known results existing in literature can be
reproduced. Furthermore, we have obtained new results, not
previously obtained, as for the case of three-point correlation
functions. The next step we have in program is the study of the
spin-$1/2$ for the 2D lattice. This step is a very hard task; the
properties of the lattice are different and the introduction of
new concepts for writing down the self-consistent equations for
the unknown parameters will be necessary.

\section{The model}

Let us consider the spin-$\frac{1}{2}$ Ising model with
nearest-neighbor interactions, in presence of an uniform external
magnetic field $h$, on a Cayley tree with coordination number $z$.
The Hamiltonian can be written as:
\begin{equation}
H=-hS\left( 0\right) +\sum_{p=1}^{z}H^{\left( p\right) }
\label{1}
\end{equation}
where $S\left( 0\right) $ is the spin operator at the central site
$\left(
0\right) $. The spin variables $S$ take only two values: $S=\pm 1$. $%
H^{\left( p\right) }$ is the Hamiltonian of the $p$-th sub-tree
rooted at the site $\left( 0\right) $ and can be written as:
\begin{equation}
H^{\left( p\right) }=-hS\left( p\right) -JS\left( 0\right) S\left(
p\right) +\sum_{m=1}^{z-1}H^{\left( p,m\right) }  \label{2}
\end{equation}
where $\left( p\right) $, ($p=1,...,z$) are the nearest neighbors
of $\left( 0\right) $, also termed the first shell. In turn
$H^{\left( p,m\right) }$ describes the $m$-th sub-tree rooted at
the site $\left( p\right) $. The process may be continued until we
eventually reach the boundary sites, described by the Hamiltonian:
\begin{equation}
H^{\left( p_{1},...,p_{r}\right) }=-hS\left(
p_{1},...,p_{r}\right) -JS\left( p_{1},...,p_{r-1}\right) S\left(
p_{1},...,p_{r}\right) , \label{4}
\end{equation}
where $\left( p_{1},p_{2},...,p_{r}\right) $ $\left[ p_{1}=1,...,z;\text{ }%
p_{2},p_{3},...=1,...,z-1\right] $ are the boundary points belonging to the $%
r$-th shell. In what follows we focus only on the sites deep in
the interior of the tree, so ignoring the boundary, i. e. we
concentrate on the Bethe lattice. Let us now consider the
transformation:
\begin{equation}
S\left( i\right) =2n\left( i\right) -1  \label{5}
\end{equation}
where $i$ is a generic site of the lattice,
\begin{equation}
n\left( i\right) =c^{\dagger }\left( i\right) c\left( i\right)
\label{6}
\end{equation}
is the density operator for a spinless fermionic field, $c\left(
i\right) $ and $c^{\dagger }\left( i\right) $ being the
annihilation and creation operators satisfying the canonical
anti-commutation relations:
\begin{equation}
\begin{array}{c}
\left\{ c\left( \mathbf{i},t\right) ,c^{\dagger }\left(
\mathbf{j},t\right)
\right\} =\delta _{\mathbf{ij}} \\
\left\{ c\left( \mathbf{i},t\right) ,c\left( \mathbf{j},t\right)
\right\}
=\left\{ c^{\dagger }\left( \mathbf{i},t\right) ,c^{\dagger }\left( \mathbf{j%
},t\right) \right\} =0
\end{array}
.  \label{7}
\end{equation}

In this way a mapping is established between the
spin-$\frac{1}{2}$ Ising model and a model of Fermi particles with
intersite Coulomb interactions on the Bethe lattice, where the
correspondence between the Ising and the fermionic variables is:
\begin{equation}
\begin{array}{ccc}
S=1 & \Rightarrow & n=1 \\
S=-1 & \Rightarrow & n=0
\end{array}
.  \label{8}
\end{equation}
The Ising Hamiltonian, eqs. (\ref{1})-(\ref{4}), with the transformation (%
\ref{5}), takes the form:
\begin{equation}
\begin{array}{c}
H=E_{0}+2\left( zJ-h\right) n\left( 0\right) +\sum_{p=1}^{z}\widehat{H}%
^{\left( p\right) } \\
\widehat{H}^{\left( p\right) }=2\left( zJ-h\right) n\left(
p\right) -4Jn\left( 0\right) n\left( p\right)
+\sum_{m=1}^{z-1}\widehat{H}^{\left(
p,m\right) } \\
\vdots \\
\widehat{H}^{\left( p_{1},...,p_{r}\right) }=2\left( zJ-h\right)
n\left( p_{1},...,p_{r}\right) -4Jn\left( p_{1},...,p_{r-1}\right)
n\left( p_{1},...,p_{r}\right)
\end{array}
,  \label{9}
\end{equation}
where the constant term $E_{0}$ is defined as:
\begin{equation}
E_{0}=h+z\left( h-J\right) \sum_{p=1}^{r}\left( z-1\right)
^{p-1}=h+z\left( h-J\right) \frac{\left( z-1\right) ^{r}-1}{z-2}.
\label{10}
\end{equation}
We immediately recognize the chemical potential $\mu =2\left(
h-zJ\right) $ and the potential strength $V=-4J$ in a fermionic
language. Also here we ignore the boundary sites and reduce to the
Bethe lattice. Such an Hamiltonian enjoys the particle-hole
symmetry, that is, it turns out to be invariant under the
transformation $n\rightarrow 1-n$, which in the spin
language corresponds to the spin-inversion symmetry $S\rightarrow -S$, $%
h\rightarrow -h$; in particular the chemical potential as a
function of $n$ scales as
\begin{equation}
\mu \left( 1-n\right) =zV-\mu \left( n\right) .  \label{cps}
\end{equation}
This scaling law implies that the magnetization vanishes in zero
external
magnetic field. However, as it will be shown in Section 7 the Hamiltonian (%
\ref{1}) and/or (\ref{9}) admits also solutions exhibiting a
spontaneous breakdown of the particle-hole symmetry; that is a
magnetization different from zero in absence of magnetic field.

We see that the density operator satisfies the equation of motion:
\begin{equation}
\mathrm{i}\frac{\partial }{\partial t}n\left( i\right) =\left[
n\left( i\right) ,H\right] =0,  \label{11}
\end{equation}
so that standard methods based on the use of equations of motion
and Green's function formalism are not immediately applicable in
terms of this operator. The relevant equation of motion to be
considered is:
\begin{equation}
\mathrm{i}\frac{\partial }{\partial t}c\left( i\right) =-\mu
c\left( i\right) -4zJc\left( i\right) n^{\alpha }\left( i\right)
\label{12}
\end{equation}
where
\begin{equation}
n^{\alpha }\left( i\right) =\frac{1}{z}\sum_{p=1}^{z}n\left(
i,p\right) , \label{13}
\end{equation}
$\left( i,p\right) $ being the nearest neighbors of the site $i$.

In the next Section, we will show in detail how to deal with such
an issue and build up the formalism. We shall put the attention to
the fermionic system and will solve the Hamiltonian (\ref{9}) by
using the formalism of the equation of motion and Green's function
method \cite
{professore1,professore2,professore3,professore4,professore5,professore6}.

\section{Composite operators and equations of motion}

In this Section, we exactly solve the Hamiltonian (\ref{9})
starting from the identification of a suitable operatorial basis
\cite {professore5,professore6}. In order to pursue this task, we
focus on the central site $\left( 0\right) $, even though we could
have chosen any other site thanks to the symmetry of the Bethe
lattice. Let us consider the following series of composite field
operators
\begin{equation}
\begin{array}{cc}
\psi _{k}\left( 0\right) =c\left( 0\right) \left[ n^{\alpha
}\left( 0\right) \right] ^{k-1}, & k=1,2,...
\end{array}
\label{14}
\end{equation}
where, according to the definition (\ref{13}) $n^{\alpha }\left( 0\right) =%
\frac{1}{z}\sum_{p=1}^{z}n\left( p\right) $, $\left( p\right) $
being the first neighbors of the site $\left( 0\right) $. By using
(\ref{11}) and (\ref {12}) it is easy to see that these operators
satisfy the hierarchy of equations of motion:
\begin{equation}
\mathrm{i}\frac{\partial }{\partial t}\psi _{k}\left( 0\right)
=\left[ \psi _{k}\left( 0\right) ,H\right] =-\mu \psi _{k}\left(
0\right) -4zJ\psi _{k+1}\left( 0\right) .  \label{16}
\end{equation}
However, we observe that the number operator $n\left( i\right) $
satisfies the following algebra:
\begin{equation}
\begin{array}{cc}
\left[ n\left( i\right) \right] ^{k}=\left[ c^{\dagger }\left(
i\right) c\left( i\right) \right] ^{k}=n\left( i\right) , &
k=1,2,...
\end{array}
.  \label{17}
\end{equation}
As shown in Appendices A and B, this algebraic property allows us
to establish the following fundamental property of the fields
$\left[ n^{\alpha }\left( i\right) \right] ^{k}$
\begin{equation}
\left[ n^{\alpha }\left( i\right) \right]
^{k}=\sum_{m=1}^{z}A_{m}^{\left( k\right) }\left[ n^{\alpha
}\left( i\right) \right] ^{m}  \label{18}
\end{equation}
where the coefficients $A_{m}^{\left( k\right) }$ are rational
numbers which can be calculated according to the scheme given in
Appendix B. Therefore, for $k=z+1$ the hierarchy of equations of
motion (\ref{16}) closes as the additional operator $\psi
_{z+2}\left( 0\right) =c\left( 0\right) \left[ n^{\alpha }\left(
0\right) \right] ^{z+1}$ can be rewritten in terms of the previous
$z+1$ elements of (\ref{14}) through the relation (\ref{18}). We
are thus able to derive a closed set of eigenoperators of the
Hamiltonian by defining the following composite operator:
\begin{equation}
\psi \left( 0\right) =\left(
\begin{array}{c}
\psi _{1}\left( 0\right) \\
\psi _{2}\left( 0\right) \\
\vdots \\
\psi _{z+1}\left( 0\right)
\end{array}
\right) =\left(
\begin{array}{c}
c\left( 0\right) \\
c\left( 0\right) n^{\alpha }\left( 0\right) \\
\vdots \\
c\left( 0\right) \left[ n^{\alpha }\left( 0\right) \right] ^{z}
\end{array}
\right)  \label{19}
\end{equation}
which satisfies the equation of motion:
\begin{equation}
\mathrm{i}\frac{\partial }{\partial t}\psi \left( 0\right) =\left[
\psi \left( 0\right) ,H\right] =\varepsilon \psi \left( 0\right) ,
\label{20}
\end{equation}
where the $\left( z+1\right) \times \left( z+1\right) $ energy matrix $%
\varepsilon $ is defined as:
\begin{equation}
\varepsilon =\left(
\begin{array}{ccccccc}
-\mu & -4zJ & 0 & \cdots & 0 & 0 & 0 \\
0 & -\mu & -4zJ & \cdots & 0 & 0 & 0 \\
0 & 0 & -\mu & \cdots & 0 & 0 & 0 \\
\vdots & \vdots & \vdots & \ddots & \vdots & \vdots & \vdots \\
0 & 0 & 0 & \cdots & -\mu & -4zJ & 0 \\
0 & 0 & 0 & \cdots & 0 & -\mu & -4zJ \\
0 & -4zJA_{1}^{\left( z+1\right) } & -4zJA_{2}^{\left( z+1\right)
} & \cdots & -4zJA_{z-2}^{\left( z+1\right) } &
-4zJA_{z-1}^{\left( z+1\right) } & -\mu -4zJA_{z}^{\left(
z+1\right) }
\end{array}
\right) .  \label{21}
\end{equation}
The eigenvalues $E_{n}$ of the energy matrix have the expressions
\begin{equation}
\begin{array}{cc}
E_{n}=-\mu -4\left( n-1\right) J, & n=1,2,...,z+1
\end{array}
.  \label{22}
\end{equation}
At this stage we can say that we have formally, but exactly,
solved Hamiltonian (\ref{9}) or its spin counterpart
(\ref{1})-(\ref{4}) as we have found for them a complete set of
eigenoperators and eigenvalues for any coordination number $z$ of
the underlying Bethe lattice. The solution is formal as we have to
compute still the correlation functions.

In order to do this, let us now define the thermal retarded
Green's function
\begin{equation}
G^{R}\left( t-t^{\prime }\right) =\left\langle R\left[ \psi \left(
0,t\right) \psi ^{\dagger }\left( 0,t^{\prime }\right) \right]
\right\rangle =\theta \left( t-t^{\prime }\right) \left\langle
\left\{ \psi \left( 0,t\right) ,\psi ^{\dagger }\left( 0,t^{\prime
}\right) \right\} \right\rangle  \label{23}
\end{equation}
where $\left\langle ...\right\rangle $ denotes the
quantum-statistical average over the grand canonical ensamble. By
introducing the Fourier transform:
\begin{equation}
G^{R}\left( t-t^{\prime }\right) =\frac{\mathrm{i}}{\left( 2\pi \right) }%
\int_{-\infty }^{+\infty }d\omega e^{-\mathrm{i}\omega \left(
t-t^{\prime }\right) }G^{R}\left( \omega \right)  \label{24}
\end{equation}
and by means of the Heisenberg equation (\ref{20}) we get the
equation:
\begin{equation}
\left[ \omega -\varepsilon \right] G^{R}\left( \omega \right) =I
\label{25}
\end{equation}
where $I$ is the normalization matrix, defined as:
\begin{equation}
I=\left\langle \left\{ \psi \left( 0,t\right) ,\psi ^{\dagger
}\left( 0,t\right) \right\} \right\rangle .  \label{26}
\end{equation}
The solution of Eq. (\ref{25}) is \cite{professore5,professore6}
\begin{equation}
G^{R}\left( \omega \right) =\sum_{n=1}^{z+1}\frac{\sigma ^{\left( n\right) }%
}{\omega -E_{n}+i\delta }  \label{27}
\end{equation}
where $\sigma ^{\left( n\right) }$ are the spectral density
matrices, to be calculated through the formula
\cite{professore5,professore6}:
\begin{equation}
\sigma _{ab}^{\left( n\right) }=\Omega _{an}\sum_{c=1}^{z+1}\left[
\Omega _{nc}\right] ^{-1}I_{cb},  \label{28}
\end{equation}
where $\Omega $ is the $\left( z+1\right) \times \left( z+1\right)
$ matrix whose columns are the eigenvectors of the matrix
$\varepsilon $.

The matrix $\Omega $ has the expression
\begin{equation}
\Omega =\left(
\begin{array}{ccccccc}
1 & z^{z} & \left( \frac{z}{2}\right) ^{z} & \cdots & \left( \frac{z}{z-2}%
\right) ^{z} & \left( \frac{z}{z-1}\right) ^{z} & 1 \\
0 & z^{z-1} & \left( \frac{z}{2}\right) ^{z-1} & \cdots & \left( \frac{z}{z-2%
}\right) ^{z-1} & \left( \frac{z}{z-1}\right) ^{z-1} & 1 \\
0 & z^{z-2} & \left( \frac{z}{2}\right) ^{z-2} & \cdots & \left( \frac{z}{z-2%
}\right) ^{z-2} & \left( \frac{z}{z-1}\right) ^{z-2} & 1 \\
\vdots & \vdots & \vdots & \ddots & \vdots & \vdots & \vdots \\
0 & z^{2} & \left( \frac{z}{2}\right) ^{2} & \cdots & \left( \frac{z}{z-2}%
\right) ^{2} & \left( \frac{z}{z-1}\right) ^{2} & 1 \\
0 & z^{1} & \left( \frac{z}{2}\right) ^{1} & \cdots & \left( \frac{z}{z-2}%
\right) ^{1} & \left( \frac{z}{z-1}\right) ^{1} & 1 \\
0 & z^{0} & \left( \frac{z}{2}\right) ^{0} & \cdots & \left( \frac{z}{z-2}%
\right) ^{0} & \left( \frac{z}{z-1}\right) ^{0} & 1
\end{array}
\right) ;  \label{29}
\end{equation}
in general, the matrix element $\Omega _{p,k}$ has the expression:
\begin{equation}
\Omega _{p,k}=\left\{
\begin{array}{cc}
1 & k=1,p=1 \\
0 & k=1,p\neq 1 \\
\left( \frac{z}{k-1}\right) ^{z+1-p} & k\neq 1
\end{array}
\right. .  \label{30}
\end{equation}
By means of the definition (\ref{26}) and of the recursion rule
(\ref{18}), the normalization matrix can be easily calculated and
has the expression
\begin{equation}
I=\left(
\begin{array}{ccccccc}
1 & I_{1,2} & I_{1,3} & \cdots & I_{1,z-1} & I_{1,z} & I_{1,z+1} \\
I_{1,2} & I_{1,3} & I_{1,4} & \cdots & I_{1,z} & I_{1,z+1} & I_{2,z+1} \\
I_{1,3} & I_{1,4} & I_{1,5} & \cdots & I_{1,z+1} & I_{2,z+1} & I_{3,z+1} \\
\vdots & \vdots & \vdots & \ddots & \vdots & \vdots & \vdots \\
I_{1,z-1} & I_{1,z} & I_{1,z+1} & \cdots & I_{z-3,z+1} &
I_{z-2,z+1} &
I_{z-1,z+1} \\
I_{1,z} & I_{1,z+1} & I_{2,z+1} & \cdots & I_{z-2,z+1} &
I_{z-1,z+1} &
I_{z,z+1} \\
I_{1,z+1} & I_{2,z+1} & I_{3,z+1} & \cdots & I_{z-1,z+1} &
I_{z,z+1} & I_{z+1,z+1}
\end{array}
\right)  \label{31}
\end{equation}
where the elements $I_{p,z+1}$ ($p=2,...,z+1$) are expressed as
\begin{equation}
I_{p,z+1}=\sum_{m=1}^{z}A_{m}^{\left( p+z-1\right) }I_{1,m+1}.
\label{32}
\end{equation}
Therefore we need to know only the elements $I_{1,k}$
($k=2,...,z+1$) which are given by
\begin{equation}
I_{1,k}=\left\langle \left[ n^{\alpha }\left( 0\right) \right]
^{k-1}\right\rangle =\kappa ^{\left( k-1\right) }.  \label{33}
\end{equation}
Then, the spectral density matrices $\sigma ^{\left( n\right) }$
can be easily calculated by means of Eq. (\ref{28}) once we keep
in mind that, according to the structure of the normalization
matrix $I$, Eqs. (\ref{31}) and (\ref{32}), there exist only $z+1$
independent matrix elements $\sigma _{1,k}^{\left( n\right) }$ for
each of the $z+1$ matrices $\sigma ^{\left( n\right) }$ while all
the others can be obtained as linear combinations of these latter
according to (\ref{32}). As a result we get:
\begin{equation}
\sigma ^{\left( n\right) }=\Sigma _{n}\Gamma ^{\left( n\right) }
\label{34}
\end{equation}
where $\Sigma _{n}$ are functions of the elements $I_{1,k}$ with $%
k=1,...,z+1 $ and $\Gamma ^{\left( n\right) }$ are numerical
matrices. In particular we have the following expressions:
\begin{equation}
\begin{array}{c}
\Sigma _{1}=\sum_{k=1}^{z+1}\Omega _{1,k}^{-1}I_{1,k} \\
\begin{array}{cc}
\Sigma _{p}=\left( \frac{z}{p-1}\right) ^{z}\sum_{k=2}^{z+1}\Omega
_{p,k}^{-1}I_{1,k}, & p=2,...,z+1
\end{array}
\end{array}
\label{35}
\end{equation}
and
\begin{equation}
\begin{array}{cc}
\Gamma _{1,k}^{\left( 1\right) }=\left( 1,0,0,...,0,0,0\right) &
k=1,...,z+1
\\
\Gamma _{1,k}^{\left( n\right) }=\left( \frac{n-1}{z}\right)
^{k-1}, & n=2,...,z+1
\end{array}
.  \label{36}
\end{equation}
The correlation function
\begin{equation}
C\left( t-t^{\prime }\right) =\left\langle \psi \left( 0,t\right)
\psi ^{\dagger }\left( 0,t^{\prime }\right) \right\rangle
=\frac{1}{\left( 2\pi \right) }\int_{-\infty }^{+\infty }d\omega
e^{-\mathrm{i}\omega \left( t-t^{\prime }\right) }C\left( \omega
\right)  \label{37}
\end{equation}
can be computed starting from Eq. (\ref{27}) and recalling the
relation:
\begin{equation}
C\left( \omega \right) =-\left[ 1+\tanh \left( \frac{\beta \omega }{2}%
\right) \right] \Im \left( G^{R}\left( \omega \right) \right) .
\label{38}
\end{equation}
Then we get:
\begin{eqnarray}
C\left( \omega \right) &=&\pi \sum_{n=1}^{z+1}\sigma ^{\left(
n\right)
}T_{n}\delta \left[ \omega -E_{n}\right]  \label{39} \\
C\left( t-t^{\prime }\right) &=&\frac{1}{2}\sum_{n=1}^{z+1}e^{-\mathrm{i}%
E_{n}\left( t-t^{\prime }\right) }\sigma ^{\left( n\right) }T_{n}
\label{40}
\end{eqnarray}
where
\begin{equation}
T_{n}=1+\tanh \left( \frac{\beta E_{n}}{2}\right) .  \label{41}
\end{equation}
Eqs. (\ref{27}) and (\ref{39}) are the exact solution of the
problem. Such a solution is only formal because the complete
knowledge of the retarded and correlation functions is not fully
achieved owing to the presence of the unknown static correlation
functions $\kappa ^{\left( m\right) }=\left\langle \left[
n^{\alpha }\left( 0\right) \right] ^{m}\right\rangle $
($m=1,...,z$) appearing in the normalization matrix $I$ due to the
non-canonical algebra satisfied by the composite field operator
$\psi \left( 0\right) $. Such unknown parameters will be
calculated according to the self-consistent scheme given in the
following Section.

\section{Self-consistency}

As we have shown in the previous Section, a complete solution of
the model requires the knowledge of the correlators $\kappa
^{\left( m\right)
}=\left\langle \left[ n^{\alpha }\left( 0\right) \right] ^{m}\right\rangle $%
. In order to compute these quantities, let us write the Hamiltonian (\ref{9}%
) as the sum of two commuting terms:
\begin{eqnarray}
H &=&H_{0}+H_{I}  \nonumber \\
H_{I} &=&-4Jn\left( 0\right) \sum_{p=1}^{z}n\left( p\right) .
\label{42}
\end{eqnarray}
Because $\left[ H_{0},H_{I}\right] =0$, for any operator $O$ we
can write its average as
\begin{equation}
\left\langle O\right\rangle =\frac{Tr\left\{ Oe^{-\beta H}\right\} }{%
Tr\left\{ e^{-\beta H}\right\} }=\frac{Tr\left\{ Oe^{-\beta
H_{I}}e^{-\beta
H_{0}}\right\} }{Tr\left\{ e^{-\beta H_{I}}e^{-\beta H_{0}}\right\} }=\frac{%
\left\langle Oe^{-\beta H_{I}}\right\rangle _{0}}{\left\langle
e^{-\beta H_{I}}\right\rangle _{0}}  \label{43}
\end{equation}
where $\left\langle ...\right\rangle _{0}$ is the trace with
respect to the reduced\ Hamiltonian $H_{0}$
\begin{equation}
\left\langle ...\right\rangle _{0}=\frac{Tr\left\{ ...e^{-\beta
H_{0}}\right\} }{Tr\left\{ e^{-\beta H_{0}}\right\} }.  \label{44}
\end{equation}
Let us now consider the correlation functions
$C_{1,k}=\left\langle c\left(
0\right) c^{\dagger }\left( 0\right) \left[ n^{\alpha }\left( 0\right) %
\right] ^{k-1}\right\rangle ,$ $k=1,...,z+1$. By means of Eq.
(\ref{43}) we can derive the following relation:
\begin{equation}
\begin{array}{cc}
\frac{C_{1,k}}{C_{1,1}}=\frac{\left\langle c\left( 0\right)
c^{\dagger }\left( 0\right) \left[ n^{\alpha }\left( 0\right)
\right] ^{k-1}e^{-\beta H_{I}}\right\rangle _{0}}{\left\langle
c\left( 0\right) c^{\dagger }\left( 0\right) e^{-\beta
H_{I}}\right\rangle _{0}}, & k=1,...,z+1
\end{array}
.  \label{48}
\end{equation}
Now from the Pauli principle we have the algebraic relation:
\begin{equation}
c^{\dagger }\left( i\right) n\left( i\right) =0  \label{49}
\end{equation}
which leads to the property
\begin{equation}
c^{\dagger }\left( 0\right) e^{-\beta H_{I}}=c^{\dagger }\left(
0\right) . \label{50}
\end{equation}
Then Eq. (\ref{48}) takes the form
\begin{equation}
\begin{array}{cc}
\frac{C_{1,k}}{C_{1,1}}=\frac{\left\langle c\left( 0\right)
c^{\dagger }\left( 0\right) \left[ n^{\alpha }\left( 0\right)
\right] ^{k-1}\right\rangle _{0}}{\left\langle c\left( 0\right)
c^{\dagger }\left( 0\right) \right\rangle _{0}}, & k=1,...,z+1
\end{array}
.  \label{51}
\end{equation}
Now let us observe that $H_{0}$ describes a system where the
original lattice has been reduced to the central site $\left(
0\right) $ and to $z$ sublattices, all disconnected among them and
topologically equivalent to the starting one. Therefore, in the
$H_{0}$-representation the correlation functions which connect
sites belonging to disconnected graphs can be decoupled:
\begin{equation}
\begin{array}{c}
\left\langle f\left\{ n\left( 0\right) \right\} g\left\{ n\left(
p\right) \right\} \right\rangle _{0}=\left\langle f\left\{ n\left(
0\right) \right\} \right\rangle _{0}\left\langle g\left\{ n\left(
p\right) \right\}
\right\rangle _{0} \\
\left\langle g\left\{ n\left( p\right) \right\} h\left\{ n\left(
q\right) \right\} \right\rangle _{0}=\left\langle g\left\{ n\left(
p\right) \right\} \right\rangle _{0}\left\langle h\left\{ n\left(
q\right) \right\} \right\rangle _{0}
\end{array}
.  \label{52}
\end{equation}
Here $f\left\{ n\left( 0\right) \right\} $, $g\left\{ n\left(
p\right) \right\} $ and $h\left\{ n\left( q\right) \right\} $,
with $p$ and $q$ belonging to different sublattices, are any
functions of the particle density. By means of such a property,
Eq. (\ref{51}) can be cast in the following form
\begin{equation}
\begin{array}{cc}
\frac{C_{1,k}}{C_{1,1}}=\left\langle \left[ n^{\alpha }\left(
0\right) \right] ^{k-1}\right\rangle _{0}, & k=1,...,z+1
\end{array}
.  \label{53}
\end{equation}
In Appendix C we show that
\begin{equation}
\begin{array}{cc}
\left\langle \left[ n^{\alpha }\left( 0\right) \right]
^{k}\right\rangle _{0}=F^{\left( z,k\right) }\left[ X\right] &
k=1,...,z
\end{array}
\label{54}
\end{equation}
where $F^{\left( z,k\right) }\left[ X\right] $ is a polynomial of
order $k$ in the variable $X$, defined as
\begin{equation}
X=\left\langle n^{\alpha }\left( 0\right) \right\rangle _{0}=\frac{C_{1,2}}{%
C_{1,1}},  \label{55}
\end{equation}
whose explicit expression is
\begin{equation}
\begin{array}{cc}
F^{\left( z,k\right) }\left[ X\right] =\sum_{p=1}^{z}a_{p}^{\left(
z,k\right) }X^{p}, & a_{p}^{\left( z,k\right)
}=\frac{1}{z^{k}}b_{p}^{\left( k\right) }\left(
\begin{array}{c}
z \\
p
\end{array}
\right)
\end{array}
;  \label{56}
\end{equation}
here $b_{p}^{\left( k\right) }$ are some numerical coefficients
defined in Appendix C. The previous analysis shows that all the
properties of the
system can be expressed in terms of only one parameter, $X$, defined by Eq. (%
\ref{55}). In order to determine this parameter we use the
self-consistent equation
\begin{equation}
C_{1,1}=1-\kappa ^{\left( 1\right) },  \label{58}
\end{equation}
where we required the translational invariant condition
$\left\langle n^{\alpha }\left( i\right) \right\rangle
=\left\langle n\left( i\right) \right\rangle $. In order to
exploit this equation we note that from (\ref {40}), by using the
definition (\ref{28}), we get:
\begin{equation}
I_{ab}=C_{a,b}+\sum_{m,p=1}^{z+1}\Omega _{am}\left[ \Omega
_{mp}\right] ^{-1}e^{-\beta E_{m}}C_{p,b}.  \label{61}
\end{equation}
Now, writing such equation for $I_{11}$ and $I_{12}$ and recalling that $%
I_{11}=1$, $I_{12}=\kappa ^{\left( 1\right) }$, we obtain
\begin{eqnarray}
1 &=&C_{1,1}+\sum_{m,p=1}^{z+1}\Omega _{1m}\left[ \Omega
_{mp}\right]
^{-1}e^{-\beta E_{m}}C_{p,1}  \label{63} \\
\kappa ^{\left( 1\right) } &=&C_{1,2}+\sum_{m,p=1}^{z+1}\Omega
_{1m}\left[ \Omega _{mp}\right] ^{-1}e^{-\beta E_{m}}C_{p,2}.
\label{64}
\end{eqnarray}
Putting such expressions in the right hand side of the
self-consistent equation (\ref{58}) we get
\begin{equation}
C_{1,1}=C_{1,1}-C_{1,2}+\sum_{m,p=1}^{z+1}\Omega _{1m}\left[ \Omega _{mp}%
\right] ^{-1}e^{-\beta E_{m}}\left( C_{p,1}-C_{p,2}\right) ,
\label{65}
\end{equation}
which, by using Eqs. (\ref{53}) and (\ref{55}), can be rewritten
as
\begin{equation}
1=\left( 1-X\right) +\sum_{m=1}^{z+1}e^{-\beta E_{m}}W^{\left(
z,m\right) }, \label{69}
\end{equation}
where
\begin{equation}
W^{\left( z,m\right) }=\sum_{p=1}^{z+1}\Omega _{1m}\left[ \Omega
_{mp}\right] ^{-1}\left( <\left[ n^{\alpha }\left( 0\right)
\right] ^{p-1}>_{0}-\;<\left[ n^{\alpha }\left( 0\right) \right]
^{p}>_{0}\right) .  \label{68}
\end{equation}
The result (\ref{54}) allows us to express the function $W^{\left(
z,m\right) }$ in terms of the parameter $X$; indeed it can be
shown that
\begin{equation}
\begin{array}{c}
\begin{array}{cc}
W^{\left( z,m\right) }=\left(
\begin{array}{c}
z-1 \\
m-1
\end{array}
\right) X^{m-1}\left( 1-X\right) ^{z+1-m} & m=1,...,z
\end{array}
\\
W^{\left( z,z+1\right) }=0
\end{array}
.  \label{70}
\end{equation}
Then, it is possible to write Eq. (\ref{69}) as follows:
\begin{equation}
1=\left( 1-X\right) \left[ 1+\sum_{m=1}^{z}e^{-\beta E_{m}}\left(
\begin{array}{c}
z-1 \\
m-1
\end{array}
\right) X^{m-1}\left( 1-X\right) ^{z-m}\right] .  \label{71}
\end{equation}
Recalling that $E_{m}=-\mu -4\left( m-1\right) J$, $\mu =2\left(
h-zJ\right) $ and making some algebraic manipulations such
equation takes finally the form:
\begin{equation}
X=\left( 1-X\right) e^{\beta \mu }\left[ 1+\left( e^{4\beta J}-1\right) X%
\right] ^{z-1}.  \label{72}
\end{equation}
Eq. (\ref{72}) is the main result of this Section; it allows us to
determine
the parameter $X$ in terms of the external parameters $\mu $, $T$, $J$ (or $%
h $, $T$, $J$).

Let us notice that, if we define a parameter $x$ as
\begin{equation}
x=\frac{e^{2\beta J}}{1+\left( e^{4\beta J}-1\right) X},
\label{73}
\end{equation}
then Eq. (\ref{72}) can be cast in the form
\begin{equation}
e^{2\beta h}=x^{z-1}\frac{\left( e^{2\beta J}-x\right) }{\left(
xe^{2\beta J}-1\right) }  \label{74}
\end{equation}
which coincides with the one given by Baxter (see Ref.
\cite{baxter}, p. 53). In our case the quantity $X$ has a definite
physical meaning, it is the particle density of the first shell in
the $H_{0}$-representation.

\section{Local correlation functions and related physical quantities}

The aim of this Section is to compute all the local correlation
functions by expressing them in terms of the parameter $X$
introduced in Eq. (\ref{55}). The calculation of the relevant
physical quantities, that is particle density, magnetization,
susceptibility, internal energy, specific heat and entropy per
site, then easily follows.

Let us start by recalling the results (\ref{53}), (\ref{54}) and
(\ref{56})
which allow us to write the correlation functions in terms of the parameter $%
X$ as follows
\begin{equation}
\begin{array}{cc}
C_{1,k}=C_{1,1}\sum_{p=1}^{z}a_{p}^{\left( z,k-1\right) }X^{p}, &
k=2,...,z+1
\end{array}
.  \label{75}
\end{equation}
where $C_{1,1}$, due to Eq. (\ref{50}), can be expressed as:
\begin{equation}
C_{1,1}=\frac{\left\langle c\left( 0\right) c^{\dagger }\left(
0\right) \right\rangle _{0}}{\left\langle e^{-\beta
H_{I}}\right\rangle _{0}}. \label{76}
\end{equation}
In order to compute $C_{1,1}$ let us observe that in the $H_{0}$%
-representation $c\left( 0\right) $ satisfies the equation of
motion
\begin{equation}
i\frac{\partial }{\partial t}c\left( 0\right) =-\mu c\left(
0\right) . \label{77}
\end{equation}
Then it is immediate to see that:
\begin{equation}
\begin{array}{cc}
\left\langle c\left( 0\right) c^{\dagger }\left( 0\right) \right\rangle _{0}=%
\frac{1}{e^{\beta \mu }+1}, & \left\langle n\left( 0\right)
\right\rangle _{0}=\frac{1}{e^{-\beta \mu }+1}
\end{array}
.  \label{78}
\end{equation}
In order to evaluate the quantity $\left\langle e^{-\beta
H_{I}}\right\rangle _{0}$ let us observe that, by means of the
algebraic property $\left[ n\left( i\right) \right] ^{m}=n\left(
i\right) $, we can write
\begin{equation}
e^{-\beta H_{I}}=e^{4\beta Jn\left( 0\right) \sum_{p=1}^{z}n\left(
p\right) }=\prod_{p=1}^{z}\left[ 1+An\left( 0\right) n\left(
p\right) \right] \label{79}
\end{equation}
where $A=e^{4\beta J}-1$. By using the property (\ref{52}) and by
recalling that in the Bethe lattice all sites are equivalent and
the parameter $X$ satisfies Eq. (\ref{72}), straightforward
calculations show that:
\begin{equation}
\left\langle e^{-\beta H_{I}}\right\rangle _{0}=1+\left[ \left(
1+AX\right)
^{z}-1\right] \left\langle n\left( 0\right) \right\rangle _{0}=\frac{1+AX^{2}%
}{\left( 1-X\right) \left( e^{\beta \mu }+1\right) }.  \label{80}
\end{equation}
Putting (\ref{78}) and (\ref{80}) into (\ref{76}) we finally get
\begin{equation}
C_{1,1}=\frac{1-X}{1+AX^{2}}.  \label{81}
\end{equation}
Now we are ready to calculate the particle density
\begin{equation}
n=\left\langle n\left( 0\right) \right\rangle
=1-C_{1,1}=\frac{X\left( 1+AX\right) }{1+AX^{2}},  \label{82}
\end{equation}
the magnetization
\begin{equation}
m=\left\langle S\left( 0\right) \right\rangle =2\left\langle
n\left( 0\right) \right\rangle -1=\frac{X\left( 2+AX\right)
-1}{1+AX^{2}}  \label{83}
\end{equation}
and all the correlation functions
\begin{equation}
\begin{array}{cc}
C_{1,k}=\frac{1-X}{1+AX^{2}}\sum_{p=1}^{z}a_{p}^{\left(
z,k-1\right) }X^{p}, & k=2,...,z+1
\end{array}
.  \label{84}
\end{equation}

Let us now switch to the calculation of the correlation functions
\begin{equation}
\begin{array}{c}
\kappa ^{\left( k\right) }=\left\langle \left[ n^{\alpha }\left(
0\right)
\right] ^{k}\right\rangle \\
\lambda ^{\left( k\right) }=\left\langle n\left( 0\right) \left[
n^{\alpha }\left( 0\right) \right] ^{k}\right\rangle
\end{array}
,k=1,...,z.  \label{85}
\end{equation}
According to the scheme given in Appendix D we have
\begin{eqnarray}
\kappa ^{\left( k\right) }
&=&\frac{1}{1+AX^{2}}\sum_{p=1}^{z}a_{p}^{\left(
z,k\right) }X^{p}\left[ \left( 1-X\right) +\frac{X\left( 1+A\right) ^{p}}{%
\left( 1+AX\right) ^{p-1}}\right]  \label{86} \\
\lambda ^{\left( k\right) }
&=&\frac{1}{1+AX^{2}}\sum_{p=1}^{z}a_{p}^{\left( z,k\right)
}\frac{X^{p+1}\left( 1+A\right) ^{p}}{\left( 1+AX\right) ^{p-1}}.
\label{87}
\end{eqnarray}
The susceptibility per site can be calculated by means of Eq.
(\ref{83}) and has the expression:
\begin{equation}
\chi =\frac{\partial m}{\partial h}=\frac{\beta \left(
1-m^{2}\right) \left( 1+p\right) }{1-\left( z-1\right) p}
\label{91}
\end{equation}
where we introduced the parameter $p$, defined as
\begin{equation}
p=\frac{AX\left( 1-X\right) }{1+AX}.  \label{92}
\end{equation}
This expression coincides with the one given in Refs.
\cite{izmail2} with $p$ playing the role of the ratio $\gamma
^{\left( 0\right) }$ of the eigenvalues of the second order
transfer matrix $V$.

Recalling the Hamiltonian (\ref{1}), we obtain for the internal
energy per site
\begin{equation}
E\left( T\right) =\frac{1}{N}\left\langle H\right\rangle =-J\left[
m^{2}\left( 1-p\right) +p\right] -hm  \label{95}
\end{equation}
where we used the fact that the total number of points in the
graph is \cite {baxter}
\begin{equation}
N=1+z\sum_{q=1}^{r}\left( z-1\right) ^{q-1}=1+z\frac{\left( z-1\right) ^{r}-1%
}{z-2}.  \label{97}
\end{equation}
Once $E\left( T\right) $ is known, we can directly calculate the
specific heat, the free energy and the entropy (per site) by means
of the formulas:
\begin{equation}
C=\frac{dE}{dT},  \label{101}
\end{equation}
\begin{equation}
F\left( T\right) =E\left( T^{\ast }\right) -T\int_{T^{\ast }}^{T}\frac{%
E\left( \widetilde{T}\right) -E\left( T^{\ast }\right) }{\widetilde{T}^{2}}d%
\widetilde{T},  \label{105b}
\end{equation}
\begin{equation}
S\left( T\right) =\frac{E\left( T\right) -F\left( T\right) }{T},
\label{105a}
\end{equation}
where the value of $E$ is given by (\ref{95}) and the limit
$T^{\ast }\rightarrow 0$ is understood.

\section{Non local correlation functions and related physical quantities}

In this Section we will calculate the relevant non local
correlation functions; then we focus on the spin-spin one and on
the related correlation length which we compare with the results
existing in the literature \cite {izmail2}. We will show how our
procedure allows us to evaluate also higher order non local
functions with respect to the one given in Refs. \cite {izmail2}.
Further technical details are presented in Appendix E.

\subsection{Two-point correlation functions}

Let us start by defining the correlation functions
\begin{eqnarray}
K^{\left( k\right) }\left( j\right) &=&\left\langle \left[
n^{\alpha }\left(
0\right) \right] ^{k}n\left( j\right) \right\rangle  \label{106} \\
\Lambda ^{\left( k\right) }\left( j\right) &=&\left\langle n\left(
0\right) \left[ n^{\alpha }\left( 0\right) \right] ^{k}n\left(
j\right) \right\rangle \label{107}
\end{eqnarray}
where $j$ is a site at a distance of $j$ steps from the central
site. Let us make for simplicity the choice that $j$ belongs to
the $z$-th subtree (but any subtree can be chosen) and let us
focus first on the two functions:
\begin{eqnarray}
K^{\left( 1\right) }\left( j\right) &=&\left\langle n^{\alpha
}\left(
0\right) n\left( j\right) \right\rangle  \label{108} \\
\Lambda ^{\left( 0\right) }\left( j\right) &=&\left\langle n\left(
0\right) n\left( j\right) \right\rangle .  \label{109}
\end{eqnarray}
We see that $\Lambda ^{\left( 0\right) }\left( j\right) $ is a
two-point correlation function which connects two sites which are
$j$ steps apart. Observing that $\left\langle n\left( z\right)
n\left( j\right) \right\rangle $ and $\left\langle n\left(
p\right) n\left( j\right) \right\rangle _{p\neq z}$ connect two
sites which are $j-1$ and $j+1$ steps apart, respectively, it is
immediate to see that the two correlation functions $K^{\left(
1\right) }\left( j\right) $ and $\Lambda ^{\left( 0\right) }\left(
j\right) $ are related through the following relation
\begin{equation}
K^{\left( 1\right) }\left( j\right) =\frac{1}{z}\Lambda ^{\left(
0\right) }\left( j-1\right) +\frac{z-1}{z}\Lambda ^{\left(
0\right) }\left( j+1\right) .  \label{111}
\end{equation}
Let us now study the function $\Lambda ^{\left( 0\right) }\left(
j\right) $; it is immediate to see that
\begin{equation}
\begin{array}{c}
\Lambda ^{\left( 0\right) }\left( 0\right) =\left\langle n\left(
0\right)
\right\rangle =n \\
\Lambda ^{\left( 0\right) }\left( 1\right) =\left\langle n\left(
0\right) n\left( 1\right) \right\rangle =\lambda ^{\left( 1\right)
}=n^{2}+n\left(
1-n\right) p \\
\Lambda ^{\left( 0\right) }\left( 2\right) =\left\langle n\left(
0\right) n\left( 2\right) \right\rangle =\frac{1}{z-1}\left(
z\kappa ^{\left( 2\right) }-n\right) =n^{2}+n\left( 1-n\right)
p^{2}
\end{array}
\label{112}
\end{equation}
where we used the results of Appendices A and D [cfr. Eqs.
(\ref{d12})-(\ref {d14})] and we noticed that by means of
(\ref{d12}) the parameter $p$, defined by Eq. (\ref{92}), can be
expressed as $p=\frac{\lambda ^{\left( 1\right) }-n^{2}}{n\left(
1-n\right) }$. On the other hand, in Appendix E we prove the
following recursion relation
\begin{equation}
G\left( j+1\right) -pG\left( j\right) =\frac{1}{p\left( z-1\right)
}\left[ G\left( j\right) -pG\left( j-1\right) \right]  \label{116}
\end{equation}
where we defined
\begin{equation}
G\left( j\right) =\Lambda ^{\left( 0\right) }\left( j\right)
-n^{2}. \label{117}
\end{equation}
Then, the two-point density correlation functions $\Lambda
^{\left( 0\right) }\left( j\right) $ for any $j$ take the
expression
\begin{equation}
\Lambda ^{\left( 0\right) }\left( j\right) =n^{2}+n\left(
1-n\right)
p^{j}\Rightarrow \frac{\Lambda ^{\left( 0\right) }\left( j\right) -n^{2}}{%
n\left( 1-n\right) }=p^{j}.  \label{119}
\end{equation}
We are now in the position to calculate higher order correlation
functions. By putting (\ref{119}) into (\ref{111}) we get
\begin{equation}
K^{\left( 1\right) }\left( j\right) =n^{2}+\frac{1}{z}n\left(
1-n\right) \left[ p^{j-1}+\left( z-1\right) p^{j+1}\right] ,
\label{120}
\end{equation}
while, by putting the result (\ref{119}) into Eq. (\ref{e19}) of
Appendix E we obtain:
\begin{equation}
\Lambda ^{\left( 1\right) }\left( j\right) =n\lambda ^{\left( 1\right) }+%
\frac{\left( 1-n\right) n}{z}\left\{ np^{j-1}+\left[ 1+n\left(
z-2\right) \right] p^{j}+\left( 1-n\right) \left( z-1\right)
p^{j+1}\right\} . \label{119bis}
\end{equation}
By noting that the parameter $p$ can be written as
\begin{equation}
p=1-\frac{X}{n}=1-\frac{\left\langle n^{\alpha }\left( 0\right)
\right\rangle _{0}}{\left\langle n\left( 0\right) \right\rangle }
\label{115}
\end{equation}
we see that it is always $p<1$. Then the correlation functions
$\Lambda
^{\left( 0\right) }\left( j\right) $, $K^{\left( 1\right) }\left( j\right) $%
, $\Lambda ^{\left( 1\right) }\left( j\right) $ satisfy the
ergodic theorem:
\begin{equation}
\begin{array}{c}
\lim_{j\rightarrow \infty }\Lambda ^{\left( 0\right) }\left(
j\right) =\left\langle n\left( 0\right) \right\rangle \left\langle
n\left( j\right)
\right\rangle =n^{2} \\
\lim_{j\rightarrow \infty }K^{\left( 1\right) }\left( j\right)
=\left\langle n^{\alpha }\left( 0\right) \right\rangle
\left\langle n\left( j\right)
\right\rangle =n^{2} \\
\lim_{j\rightarrow \infty }\Lambda ^{\left( 1\right) }\left(
j\right) =\left\langle n\left( 0\right) n^{\alpha }\left( 0\right)
\right\rangle \left\langle n\left( j\right) \right\rangle
=n\lambda ^{\left( 1\right) }
\end{array}
.  \label{erth1}
\end{equation}
Recalling (\ref{5}), we can evaluate from (\ref{119}) the
spin-spin correlation function
\begin{equation}
\left\langle S\left( 0\right) S\left( j\right) \right\rangle
=m^{2}+\left( 1-m^{2}\right) p^{j}.  \label{121}
\end{equation}
This expression coincides with the result of Refs. \cite{izmail2}.
Now, by defining the correlation function:
\begin{equation}
G_{S}\left( j\right) =\left\langle S\left( 0\right) S\left(
j\right) \right\rangle -\left\langle S\left( 0\right)
\right\rangle \left\langle S\left( j\right) \right\rangle
\label{122}
\end{equation}
we obtain from (\ref{121})
\begin{equation}
G_{S}\left( j\right) =\left( 1-m^{2}\right) e^{-\frac{j}{\xi }}
\label{123}
\end{equation}
where the correlation length is defined as
\begin{equation}
\xi =\left[ \ln \left( \frac{1}{p}\right) \right] ^{-1}.
\label{124}
\end{equation}

\subsection{Three-point correlation functions}

Following the same line of reasoning which led us to the two-point
correlation functions, let us now calculate three-point
correlation functions. Let us define the general three-point
correlator as:
\begin{equation}
\begin{array}{cc}
T^{\left( k\right) }\left( j\mathbf{,}w\right) =\left\langle
c\left( 0\right) c^{\dagger }\left( 0\right) \left[ n^{\alpha
}\left( 0\right) \right] ^{k-1}n\left( j\right) n\left( w\right)
\right\rangle =M^{\left(
k-1\right) }\left( j\mathbf{,}w\right) -N^{\left( k-1\right) }\left( j%
\mathbf{,}w\right) & \left( k\geq 1\right)
\end{array}
\label{h1}
\end{equation}
where we introduce the new correlation functions
\begin{eqnarray}
M^{\left( k\right) }\left( j\mathbf{,}w\right) &=&\left\langle
\left[ n^{\alpha }\left( 0\right) \right] ^{k}n\left( j\right)
n\left( w\right)
\right\rangle  \label{h3} \\
N^{\left( k\right) }\left( j\mathbf{,}w\right) &=&\left\langle
n\left( 0\right) \left[ n^{\alpha }\left( 0\right) \right]
^{k}n\left( j\right) n\left( w\right) \right\rangle .
\end{eqnarray}
By $j$ and $w$ we denote two sites at a distance of $j$ and $w$
steps, respectively, with respect to the central site $\left(
0\right) $. Let us distinguish the two following cases: 1) $j$ and
$w$ belong to the same subtree; 2) $j$ and $w$ belong to different
subtrees.

\subsubsection*{Case 1}

$j$ and $w$ belong to the same subtree, which we take as the
$z$-subtree, but any subtree can be chosen. By means of
(\ref{43}), (\ref{49}) and (\ref {52}) and by noting that:
\begin{equation}
\frac{\left\langle c\left( 0\right) c^{\dagger }\left( 0\right)
\right\rangle _{0}}{\left\langle e^{-\beta H_{I}}\right\rangle _{0}}%
=C_{1,1}=1-n,  \label{happ}
\end{equation}
we can express $T^{\left( k\right) }\left( j\mathbf{,}w\right) $
as:
\begin{equation}
T^{\left( k\right) }\left( j\mathbf{,}w\right) =M^{\left(
k-1\right) }\left( j\mathbf{,}w\right) -N^{\left( k-1\right)
}\left( j\mathbf{,}w\right) =\left( 1-n\right) \left\langle \left[
n^{\alpha }\left( 0\right) \right] ^{k-1}n\left( j\right) n\left(
w\right) \right\rangle _{0}.  \label{h14}
\end{equation}
Let us concentrate the attention on the two functions $N^{\left(
0\right) }\left( j,w\right) =\left\langle n\left( 0\right) n\left(
j\right) n\left( w\right) \right\rangle $ and $M^{\left( 1\right)
}\left( j,w\right) =\left\langle n^{\alpha }\left( 0\right)
n\left( j\right) n\left( w\right) \right\rangle $. At first, we
note that these functions are related through the following
relation
\begin{equation}
M^{\left( 1\right) }\left( j,w\right) =\frac{1}{z}N^{\left(
0\right) }\left( j-1,w-1\right) +\frac{z-1}{z}N^{\left( 0\right)
}\left( j+1,w+1\right) . \label{h8}
\end{equation}
Next, let us study the function $N^{\left( 0\right) }\left(
j,w\right) $; by recalling the definitions (\ref{85}) it is
immediate to see that
\begin{equation}
\begin{array}{c}
N^{\left( 0\right) }\left( 0,0\right) =\left\langle n\left(
0\right) n\left( 0\right) n\left( 0\right) \right\rangle
=\left\langle n\left( 0\right)
\right\rangle =n \\
N^{\left( 0\right) }\left( 0,1\right) =N^{\left( 0\right) }\left(
1,0\right) =N^{\left( 0\right) }\left( 1,1\right) =\left\langle
n\left( 0\right)
n\left( 1\right) \right\rangle =\lambda ^{\left( 1\right) } \\
N^{\left( 0\right) }\left( 0,2\right) =N^{\left( 0\right) }\left(
2,0\right) =N^{\left( 0\right) }\left( 2,2\right) =\left\langle
n\left( 0\right) n\left( 2\right) \right\rangle
=\frac{1}{z-1}\left( z\kappa ^{\left(
2\right) }-n\right) \\
N^{\left( 0\right) }\left( 1,2\right) =N^{\left( 0\right) }\left(
2,1\right) =\left\langle n\left( 0\right) n\left( 1\right) n\left(
2\right) \right\rangle =\frac{1}{z-1}\left( z\lambda ^{\left(
2\right) }-\lambda ^{\left( 1\right) }\right)
\end{array}
.  \label{h9}
\end{equation}
Recalling the relations (\ref{d14}) in Appendix D and the expression $p=%
\frac{\lambda ^{\left( 1\right) }-n^{2}}{n\left( 1-n\right) }$,
the correlation function $N^{\left( 0\right) }\left( j,w\right) $
can be written in the closed form
\begin{equation}
\begin{array}{cc}
N^{\left( 0\right) }\left( j,w\right) =n^{3}+n^{2}\left(
1-n\right) \left( p^{j}+p^{w-j}\right) +n\left( 1-n\right)
^{2}p^{w}, & j,w=0,1,2.
\end{array}
\label{h11}
\end{equation}
In order to evaluate $N^{\left( 0\right) }\left(
j\mathbf{,}w\right) $ for all values of $j$ and $w$ we need a
recursion formula as the one in Eq. (\ref {116}), which we now
derive following the same steps outlined in Appendix E for the
function $\Lambda ^{\left( 0\right) }\left( j\right) $.

In the $H_{0}$-representation $N^{\left( 0\right) }\left( j\mathbf{,}%
w\right) $ can be written as
\begin{equation}
N^{\left( 0\right) }\left( j,w\right) =\left\langle n\left(
0\right) n\left( j\right) n\left( w\right) \right\rangle
=\frac{\left\langle n\left( 0\right)
n\left( j\right) n\left( w\right) e^{-\beta H_{I}}\right\rangle _{0}}{%
\left\langle e^{-\beta H_{I}}\right\rangle _{0}}.  \label{h16}
\end{equation}
Recalling that (cfr. (\ref{79})) $e^{-\beta
H_{I}}=\prod_{p=1}^{z}\left[ 1+An\left( 0\right) n\left( p\right)
\right] $, by making use of Eqs. (\ref {78}) and (\ref{80}), as
well as of the equation (\ref{72}) for the parameter $X$ , we have
\begin{equation}
N^{\left( 0\right) }\left( j,w\right)
=\frac{X}{1+AX^{2}}\left\langle n\left( j\right) n\left( w\right)
\left( 1+An\left( z\right) \right) \right\rangle _{0}.
\label{h18}
\end{equation}
Let us now calculate the function $N^{\left( 1\right) }\left(
j,w\right) =\left\langle n\left( 0\right) n^{\alpha }\left(
0\right) n\left( j\right) n\left( w\right) \right\rangle $. By
following the same procedure, we obtain for $j,w>1$:
\begin{eqnarray}
N^{\left( 1\right) }\left( j,w\right) &=&\frac{\left( z-1\right) }{z}\frac{%
X^{2}\left( 1+A\right) }{\left( 1+AX\right) \left( 1+AX^{2}\right) }%
\left\langle n\left( j\right) n\left( w\right) \left( 1+An\left(
z\right)
\right) \right\rangle _{0}  \nonumber \\
&&+\frac{1}{z}\frac{X\left( 1+A\right) }{1+AX^{2}}\left\langle
n\left( z\right) n\left( j\right) n\left( w\right) \right\rangle
_{0}.  \label{h20}
\end{eqnarray}
For simplicity, let us restrict the analysis to the case where $w$ follows $%
j $ (i. e. we can write $w=j+r$, where $r$ is the number of steps
necessary to go from $w$ to $j$). In this case we have $M^{\left(
0\right) }\left( j,w\right) =\Lambda ^{\left( 0\right) }\left(
r\right) $ and Eq. (\ref{h14}) for $k=1$ gives
\begin{equation}
\Lambda ^{\left( 0\right) }\left( r\right) -N^{\left( 0\right) }\left( j%
\mathbf{,}j+r\right) =\left( 1-n\right) \left\langle n\left(
j\right) n\left( j+r\right) \right\rangle _{0}.  \label{h15}
\end{equation}
By combining (\ref{h15}) and (\ref{h18}) we can express the two quantities $%
\left\langle n\left( j\right) n\left( j+r\right) \right\rangle _{0}$ and $%
\left\langle n\left( j\right) n\left( j+r\right) n\left( z\right)
\right\rangle _{0}$ in terms of $N^{\left( 0\right) }\left( j\mathbf{,}%
j+r\right) $ and $\Lambda ^{\left( 0\right) }\left( r\right) $ as
follows:
\begin{eqnarray}
\left\langle n\left( j\right) n\left( j+r\right) \right\rangle _{0} &=&\frac{%
1}{\left( 1-n\right) }\left[ \Lambda ^{\left( 0\right) }\left(
r\right)
-N^{\left( 0\right) }\left( j\mathbf{,}j+r\right) \right]  \nonumber \\
\left\langle n\left( j\right) n\left( j+r\right) n\left( z\right)
\right\rangle _{0} &=&\frac{1+AX^{2}}{AX\left( 1-X\right)
}N^{\left( 0\right) }\left( j\mathbf{,}j+r\right)
-\frac{1}{A}\frac{1}{\left( 1-n\right) }\Lambda ^{\left( 0\right)
}\left( r\right) .
\end{eqnarray}
By using the relation (\ref{h14}) for $k=2$ and by observing that
\begin{equation}
\left\langle n^{\alpha }\left( 0\right) n\left( j\right) n\left(
j+r\right) \right\rangle _{0}=\frac{1}{z}\left\langle n\left(
j\right) n\left( j+r\right) n\left( z\right) \right\rangle
_{0}+\frac{z-1}{z}X\left\langle n\left( j\right) n\left(
j+r\right) \right\rangle _{0},  \label{h23}
\end{equation}
we can express the function $M^{\left( 1\right) }\left( j\mathbf{,}%
j+r\right) $ in terms of $N^{\left( 0\right) }\left(
j\mathbf{,}j+r\right) $ and $\Lambda ^{\left( 0\right) }\left(
r\right) $ as
\begin{equation}
M^{\left( 1\right) }\left( j\mathbf{,}j+r\right) =\left[
\frac{\left(
z-1\right) p}{z}+\frac{1}{zp}\right] N^{\left( 0\right) }\left( j\mathbf{,}%
j+r\right) +\left[ \frac{\left( z-1\right)
p}{z}-\frac{1}{zp}\right] n\left( 1-p\right) \Lambda ^{\left(
0\right) }\left( r\right) .  \label{h24}
\end{equation}
Recalling that [see (\ref{119})] $\Lambda ^{\left( 0\right)
}\left( r\right) =n^{2}+n\left( 1-n\right) p^{r}$ we obtain from
(\ref{h8}) the relevant recursion rule
\begin{equation}
\widehat{N}^{\left( 0\right) }\left( j+1,w+1\right)
-p\widehat{N}^{\left(
0\right) }\left( j,w\right) =\frac{1}{p\left( z-1\right) }\left[ \widehat{N}%
^{\left( 0\right) }\left( j,w\right) -p\widehat{N}^{\left(
0\right) }\left( j-1,w-1\right) \right]  \label{hr2}
\end{equation}
where we defined
\begin{equation}
\widehat{N}^{\left( 0\right) }\left( j,w\right) =N^{\left(
0\right) }\left( j,w\right) -n^{3}-n^{2}\left( 1-n\right) p^{w-j}.
\label{hr1}
\end{equation}
By recalling the result (\ref{h11}), it is easy to see that the
following
expressions hold for any $j$ and $w$%
\begin{eqnarray}
\frac{\widehat{N}^{\left( 0\right) }\left( j,w\right) }{n\left(
1-n\right) }
&=&np^{j}+\left( 1-n\right) p^{w}  \label{hr6} \\
N^{\left( 0\right) }\left( j,w\right) &=&n^{3}+n^{2}\left(
1-n\right) \left( p^{j}+p^{w-j}\right) +n\left( 1-n\right)
^{2}p^{w}.  \label{h26}
\end{eqnarray}
By using the transformation (\ref{5}) we obtain from (\ref{h26})
the expression of the three-spin correlation function:
\begin{equation}
\left\langle S\left( 0\right) S\left( j\right) S\left( w\right)
\right\rangle =m^{3}+m\left( 1-m^{2}\right) \left(
p^{j}-p^{w}+p^{w-j}\right) ;  \label{threesc}
\end{equation}
this expression agrees with the one given in Ref. \cite{marsh},
where it was calculated for the one dimensional case (i. e.
$z=2$). It is interesting to notice that the expressions of the
spin correlation functions [cfr. (\ref {121}) and (\ref{threesc})]
depend on the coordination number $z$ only through the parameters
$p$ and $m$.

We are now in position to calculate the correlation functions
$M^{\left( 1\right) }\left( j,w\right) $ and $N^{\left( 1\right)
}\left( j,w\right) $. Straightforward calculations give
\begin{equation}
M^{\left( 1\right) }\left( j,w\right) =n^{3}+\frac{1}{z}n\left(
1-n\right) \left\{ n\left( p^{j-1}+p^{w-j}\right) +\left(
1-n\right) p^{w-1}+\left(
z-1\right) \left[ n\left( p^{j+1}+p^{w-j}\right) +\left( 1-n\right) p^{w+1}%
\right] \right\} ,  \label{h27}
\end{equation}
\begin{eqnarray}
N^{\left( 1\right) }\left( j,w\right) &=&n^{2}\lambda ^{\left(
1\right) }+n^{2}\left( 1-n\right) \left[ np^{w-j}+\left(
1-n\right) p^{w-j+1}\right] +
\nonumber \\
&&\frac{n\left( 1-n\right) }{z}\left[ n^{2}p^{j-1}+\left(
1-n\right)
^{2}p^{w}+n\left( 1-n\right) \left( p^{w-1}+p^{j}\right) \right] + \\
&&\frac{n\left( 1-n\right) \left( z-1\right) }{z}\left[
n^{2}p^{j}+\left( 1-n\right) ^{2}p^{w+1}+n\left( 1-n\right) \left(
p^{j+1}+p^{w}\right) \right] .  \nonumber
\end{eqnarray}
Let us notice that all the correlators $N^{\left( 0\right) }\left(
j,w\right) $, $M^{\left( 1\right) }\left( j,w\right) $ and
$N^{\left( 1\right) }\left( j,w\right) $ satisfy the ergodic
theorem.

\subsubsection*{Case 2}

Let us now switch to the second case: $j$ and $w$ belong to
different subtrees, which we take as the $z$- and $\left(
z-1\right) $-subtree, respectively. By performing the same steps
which led to Eq. (\ref{h14}) and
by noting that in such a case $M^{\left( 0\right) }\left( j\mathbf{,}%
w\right) =\Lambda ^{\left( 0\right) }\left( j+w\right) $ we obtain
the relation
\begin{equation}
N^{\left( 0\right) }\left( j\mathbf{,}w\right) =\Lambda ^{\left(
0\right) }\left( j+w\right) -\left( 1-n\right) \left\langle
n\left( j\right) \right\rangle _{0}\left\langle n\left( w\right)
\right\rangle _{0}. \label{m10}
\end{equation}
Recalling the expressions (\ref{e14}) for $\left\langle n\left(
j\right) \right\rangle _{0}$ and (\ref{119}) for $\Lambda ^{\left(
0\right) }\left( j+w\right) $, we obtain for any $j$ and $w$
\begin{equation}
N^{\left( 0\right) }\left( j,w\right) =n^{3}+n^{2}\left(
1-n\right) \left( p^{j}+p^{w}\right) +n\left( 1-n\right)
^{2}p^{j+w}.  \label{m13}
\end{equation}
To calculate higher order correlation functions we observe that
\begin{equation}
M^{\left( 1\right) }\left( j,w\right) =\frac{1}{z}N^{\left(
0\right) }\left(
j-1,w+1\right) +\frac{1}{z}N^{\left( 0\right) }\left( j+1,w-1\right) +\frac{%
z-2}{z}N^{\left( 0\right) }\left( j+1,w+1\right) .  \label{m8}
\end{equation}
Putting (\ref{m13}) into (\ref{m8}) we have
\begin{eqnarray}
M^{\left( 1\right) }\left( j,w\right) &=&n^{3}+\frac{1}{z}n\left(
1-n\right)
[n\left( p^{j-1}+p^{w-1}\right) +2\left( 1-n\right) p^{j+w}+  \nonumber \\
&&\left( z-1\right) n\left( p^{j+1}+p^{w+1}\right) +\left(
z-2\right) \left( 1-n\right) p^{j+w+2}].  \label{m14}
\end{eqnarray}
In order to calculate $N^{\left( 1\right) }\left( j,w\right) $ let
us observe that
\begin{equation}
N^{\left( 1\right) }\left( j,w\right) =M^{\left( 1\right) }\left(
j,w\right) -T^{\left( 2\right) }\left( j\mathbf{,}w\right)
=M^{\left( 1\right) }\left( j,w\right) -\left( 1-n\right)
\left\langle n^{\alpha }\left( 0\right) n\left( j\right) n\left(
w\right) \right\rangle _{0}.  \label{m15}
\end{equation}
which, by noting that
\begin{equation}
\left\langle n^{\alpha }\left( 0\right) n\left( j\right) n\left(
w\right) \right\rangle _{0}=\frac{1}{z}\left\langle n\left(
z\right) n\left( j\right)
\right\rangle _{0}\left\langle n\left( w\right) \right\rangle _{0}+\frac{1}{z%
}\left\langle n\left( z-1\right) n\left( w\right) \right\rangle
_{0}\left\langle n\left( j\right) \right\rangle _{0}+\frac{z-2}{z}%
X\left\langle n\left( j\right) \right\rangle _{0}\left\langle
n\left( w\right) \right\rangle _{0},  \label{m16}
\end{equation}
and recalling (\ref{e14}) and (\ref{m14}), becomes
\begin{eqnarray}
N^{\left( 1\right) }\left( j,w\right) &=&n^{2}\lambda ^{\left( 1\right) }+%
\frac{\left[ \left( z-2\right) n+1\right] }{z}n^{2}\left(
1-n\right) \left( p^{j}+p^{w}\right) +\frac{\left( z-1\right)
}{z}n^{2}\left( 1-n\right)
^{2}\left( p^{j+1}+p^{w+1}\right)  \nonumber \\
&&+\frac{n^{3}\left( 1-n\right) }{z}\left( p^{j-1}+p^{w-1}\right) +\frac{2}{z%
}n^{2}\left( 1-n\right) ^{2}p^{j+w-1}+\frac{\left( z-2\right) }{z}%
n^{3}\left( 1-n\right) p^{j+w+1}  \nonumber \\
&&+\frac{\left[ 2\left( 1-2n\right) -\left( z-4\right) n^{2}\right] }{z}%
n\left( 1-n\right) p^{j+w}+\frac{\left( z-2\right) }{z}n\left(
1-n\right) ^{2}p^{j+w+2}  \label{m17}
\end{eqnarray}
for any $j$ and $w$. Also in this case the ergodic theorem is
satisfied.

\section{Results}

In the previous Sections we have shown that all the properties of
the system are expressed in terms of the correlator
$X=\left\langle n^{\alpha }\left( 0\right) \right\rangle _{0}$.
This quantity is determined in terms of the external parameters
$J$, $T$, $h$ by solving the equation (\ref{72}), which is a
polynomial of order $z$ in the variable $X$. In this Section we
discuss the solutions of Eq. (\ref{72}) and present the results
obtained for various properties: the magnetization, the
susceptibility, the specific heat, the free energy and the
entropy. We shall discuss separately the cases of zero and finite
magnetic field, by focusing the analysis to a ferromagnetic
coupling (i. e. $J>0$).

\subsection{Zero magnetic field}

In the case of zero magnetic field we have $\mu =-2zJ$; then, it
is useful to define $K=e^{2\beta J}$ so that the equation
(\ref{72}) takes the form
\begin{equation}
XK^{z}=\left( 1-X\right) \left[ 1+\left( K^{2}-1\right) X\right]
^{z-1}. \label{126}
\end{equation}
It is easy to see that
\begin{equation}
X=\frac{1}{1+K}=\frac{1}{e^{2\beta J}+1}  \label{127}
\end{equation}
is always a solution of the equation (\ref{126}) for any value of
the
coordination number $z$. By putting (\ref{127}) into (\ref{82}) and (\ref{83}%
) we have
\begin{equation}
\begin{array}{cc}
n=\left\langle n\left( 0\right) \right\rangle =\frac{1}{2}, &
m=\left\langle S\left( 0\right) \right\rangle =0
\end{array}
.  \label{128}
\end{equation}
The particle density and the magnetization do not depend on the
temperature and on the coordination number $z$. This is a
manifestation of the particle-hole symmetry, when we recall the
scaling law (\ref{cps}) for the chemical potential. But
(\ref{126}) may admit other solutions which break the
particle-hole symmetry. In particular, let us study if there is a
critical temperature $T_{c}$ such that the magnetization is
different from zero for $T<T_{c}$. In order to determine $T_{c}$
let us expand (\ref{126}) in power series of $X$ around the point
given by (\ref{127}). At first order we obtain
\begin{equation}
\left( X-\frac{1}{1+K}\right) \left[ z\left( K-1\right) -2K\right]
=0. \label{129}
\end{equation}
Therefore, besides the solution (\ref{127}) there may be other
solutions when $K=\frac{z}{z-2}$. Such an equation shows that
there is a critical temperature $T_{c}$, given by
\begin{equation}
2J=k_{B}T_{c}\ln \left( \frac{z}{z-2}\right) ,  \label{131}
\end{equation}
such that for $T<T_{c}$ we may have solutions which spontaneously
break the particle-hole symmetry and exhibit a magnetization
different from zero. Let
us notice that the case $z=2$ (i.e. the one-dimensional chain) gives $%
T_{c}=0 $. Let us also point out that Eq. (\ref{131}) admits a
solution only when $J>0$. For negative $J$ (i.e. antiferromagnetic
coupling) there is no solution. If $z$ is even, the equation
(\ref{126}) admits another solution
\begin{equation}
X=\frac{1}{1-K}=\frac{1}{1-e^{2\beta J}}  \label{132}
\end{equation}
which also gives the results quoted in (\ref{128}). However, such
a solution describes an unstable system: the energy is a
decreasing function of temperature and the parameter $p$ is larger
than one. This solution will be disregarded in the following.

Generally, for $z>2$ we have the following situation:

\begin{itemize}
\item  $z$ even
\[
\begin{array}{c}
T<T_{c}\left\{
\begin{array}{c}
\text{2 solutions corresponding to }n=\frac{1}{2}\text{ and }m=0\text{ } \\
\text{2 solutions corresponding to }\pm m\neq 0 \\
\text{the remaining roots are complex}
\end{array}
\right. \\
T>T_{c}\left\{
\begin{array}{c}
\text{2 solutions corresponding to }n=\frac{1}{2}\text{ and }m=0 \\
\text{the remaining roots are complex}
\end{array}
\right.
\end{array}
\]

\item  $z$ odd
\[
\begin{array}{c}
T<T_{c}\left\{
\begin{array}{c}
\text{1 solution corresponding to }n=\frac{1}{2}\text{ and }m=0\text{ } \\
\text{2 solutions corresponding to }\pm m\neq 0 \\
\text{the remaining roots are complex}
\end{array}
\right. \\
T>T_{c}\left\{
\begin{array}{c}
\text{1 solution corresponding to }n=\frac{1}{2}\text{ and }m=0 \\
\text{the remaining roots are complex}
\end{array}
\right.
\end{array}
\]
\end{itemize}

By considering the following items: (i) the broken symmetry
solution (i. e., $m\neq 0$) has a free energy lower than the one
corresponding to the symmetric solution, (ii) the solution
(\ref{132}) is disregarded because not physical, (iii) the two
solutions corresponding to $\pm m$ are physically equivalent, (iv)
all the complex solutions are disregarded, we can assert that the
equation (\ref{126}) admits only one solution of physical
interest.

For $T>T_{c}$ we have the following results:
\begin{equation}
\begin{array}{cccc}
X=\frac{1}{2}\left[ 1-\tanh \left( \beta J\right) \right] &
n=\frac{1}{2} & \lambda ^{\left( 1\right) }=\frac{1}{4}\left[
1+\tanh \left( \beta J\right)
\right] & p=\tanh \left( \beta J\right) \\
m=0 & E=-J\tanh \left( \beta J\right) & C=k_{B}\left[ \beta J\sec \mathrm{h}%
\left( \beta J\right) \right] ^{2} & \chi =\frac{\beta \left[
1+\tanh \left( \beta J\right) \right] }{1-\left( z-1\right) \tanh
\left( \beta J\right) }
\end{array}
.  \label{tgg}
\end{equation}

For $T<T_{c}$ the breaking symmetry solution depends on $z$. We
shall present results for $z=3$ and $z=4$. For the case of $z=3$
the critical
temperature is given by $\frac{k_{B}T_{c}}{J}=\frac{2}{\ln 3}\approx 1.82048$%
. The solution of (\ref{126}) is:
\begin{equation}
\begin{array}{cc}
X=\frac{\left( K+1\right) \left( K-2\right) +K\sqrt{\left(
K+1\right) \left(
K-3\right) }}{2\left( K^{2}-1\right) } &  \\
n=\frac{\left( K+1\right) \left( K-2\right) +K\sqrt{\left(
K+1\right) \left( K-3\right) }}{2\left( K+1\right) \left(
K-2\right) } & m=\frac{K\sqrt{\left(
K+1\right) \left( K-3\right) }}{\left( K+1\right) \left( K-2\right) } \\
p=\frac{1}{K-1} & \chi =\frac{4\beta K}{\left( K-3\right) \left(
K-2\right) ^{2}\left( K+1\right) }
\end{array}
.  \label{z3sol}
\end{equation}
For the case of $z=4$ the critical temperature is given by $\frac{k_{B}T_{c}%
}{J}=\frac{2}{\ln 2}\approx 2.88539$. The solution of (\ref{126})
is:
\begin{equation}
\begin{array}{cc}
X=\frac{K^{2}-2+K\sqrt{K^{2}-4}}{2\left( K^{2}-1\right) } &  \\
n=\frac{K^{2}-2+K\sqrt{K^{2}-4}}{2\left( K^{2}-2\right) } & m=\frac{K\sqrt{%
K^{2}-4}}{K^{2}-2} \\
p=\frac{1}{K^{2}-1} & \chi =\frac{4\beta K^{2}}{\left(
K^{2}-4\right) \left( K^{2}-2\right) ^{2}}
\end{array}
.  \label{z4sol}
\end{equation}
For all values of $z$ the parameter $\lambda ^{\left( 1\right) }$
and the internal energy $E$ can be calculated by means of the
expressions
\begin{equation}
\begin{array}{cc}
\lambda ^{\left( 1\right) }=n\left[ p+n\left( 1-p\right) \right] ,
& E=4Jn\left( 1-n\right) \left( 1-p\right) -2J
\end{array}
.  \label{intel}
\end{equation}

In Fig. 1 we plot the magnetization per site $m=\left\langle
S\left( 0\right) \right\rangle $ as a function of the temperature,
expressed in units of $J$, for the values of the coordination
number $z=3$ and $z=4$. As expected, the magnetization decreases
by increasing $T$ and vanishes at the critical temperature
$T_{c}$, determined by (\ref{131}). By expanding the parameter
$K=e^{2\beta J}$ around the critical temperature $T_{c}$:
\begin{equation}
K=K_{c}\left[ 1+bt\right] +O\left( t^{2}\right)  \label{kappaex}
\end{equation}
where
\begin{equation}
\begin{array}{ccc}
t=\frac{T_{c}-T}{T_{c}}, & K_{c}=\frac{z}{z-2}, & b=\ln \left( \frac{z}{z-2}%
\right)
\end{array}
,  \label{valuesex}
\end{equation}
we can easily show that close to $T_{c}$ the magnetization behaves
as
\begin{equation}
m=\left\{
\begin{array}{cc}
\frac{3\sqrt{3bt}}{2}+O\left( t^{\frac{3}{2}}\right) & z=3 \\
2\sqrt{2bt}+O\left( t^{\frac{3}{2}}\right) & z=4
\end{array}
\right.  \label{mag1}
\end{equation}
with critical exponents $\beta =\frac{1}{2}$, in agreement with
Refs. \cite {baxter,izmail2}. The behaviour of the parameter
$p=1-\frac{\left\langle n^{\alpha }\left( 0\right) \right\rangle
_{0}}{\left\langle n\left( 0\right) \right\rangle }$ as a function
of $\frac{T}{J}$ is shown in Fig. 2. By
increasing the temperature, $p$ first increases up to the maximum value $%
\left( p\right) _{T=T_{c}}=\frac{1}{z-1}$, then decreases. It is always $p<1$%
: this condition implies the ergodic behaviour of the spin
correlation
functions, when we recall the results of Section 6. We notice that for $%
T>T_{c}$ the value of $p$ is the same for all values of $z$.

\begin{figure}[tbph]
\centering\includegraphics*[width=0.5\linewidth]{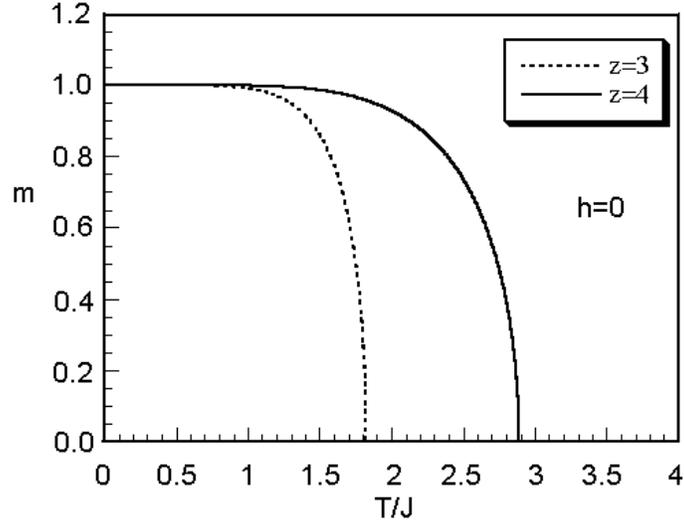}
\caption{The magnetization $m$ is plotted against $T/J$ for $z=3$
and $z=4$ and zero magnetic field.} \label{figura1}
\end{figure}

\begin{figure}[tbph]
\centering\includegraphics*[width=0.5\linewidth]{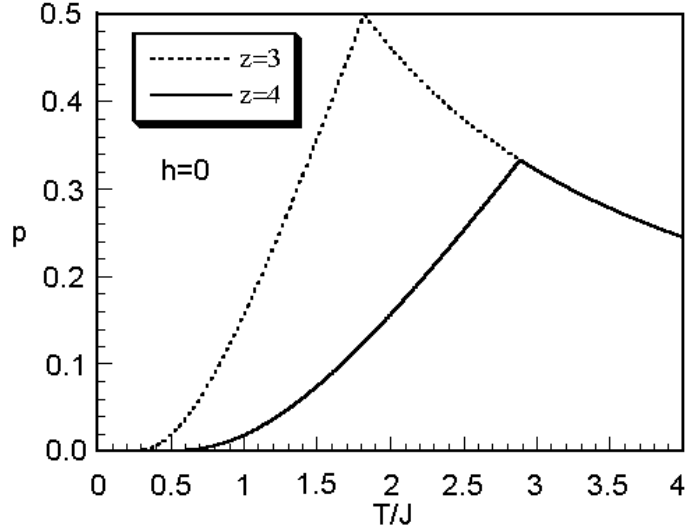}
\caption{The parameter $p$ is plotted against $T/J$ for $z=3$ and
$z=4$ and zero magnetic field.} \label{figura2}
\end{figure}

In Fig. 3 we report the temperature dependence of the spin
susceptibility per site $\chi =\left( \frac{\partial m}{\partial
h}\right) _{h=0}$. This quantity diverges at $T=T_{c}$ with
critical exponents $\gamma =\gamma ^{\prime }=-1$:
\begin{equation}
\chi =\left\{
\begin{array}{cc}
\frac{\left( T_{c}-T\right) ^{-1}}{b\left( z-2\right) } & T<T_{c} \\
\frac{2\left( T-T_{c}\right) ^{-1}}{b\left( z-2\right) } & T>T_{c}
\end{array}
\right. .  \label{susc1}
\end{equation}
To the contrary of $p$, above $T_{c}$ the susceptibility changes
with $z$, as can be seen by Eq. (\ref{91}). The specific heat per
site is reported in Fig. 4 as a function of the temperature. We
observe a jump in correspondence of $T_{c}$, with critical
exponents $\alpha =0$, as expected for a second order phase
transition. It can be shown that the jump $\Delta C$ at $T=T_{c}$
is given by:
\begin{equation}
\Delta C=\left\{
\begin{array}{cc}
\frac{21J^{2}}{4T_{c}^{2}}-\frac{4J^{2}K_{c}}{\left(
1+K_{c}\right)
^{2}T_{c}^{2}}=\frac{9}{8}\left[ \ln 3\right] ^{2} & z=3 \\
\frac{80J^{2}}{9T_{c}^{2}}-\frac{4J^{2}K_{c}}{\left(
1+K_{c}\right) ^{2}T_{c}^{2}}=2\left[ \ln 2\right] ^{2} & z=4
\end{array}
\right. .  \label{csjump}
\end{equation}
We notice that the jump decreases with $z$, and that above $T_{c}$
the behaviour of the specific heat does not depend on the value of
$z$.

\begin{figure}[tbph]
\centering\includegraphics*[width=0.5\linewidth]{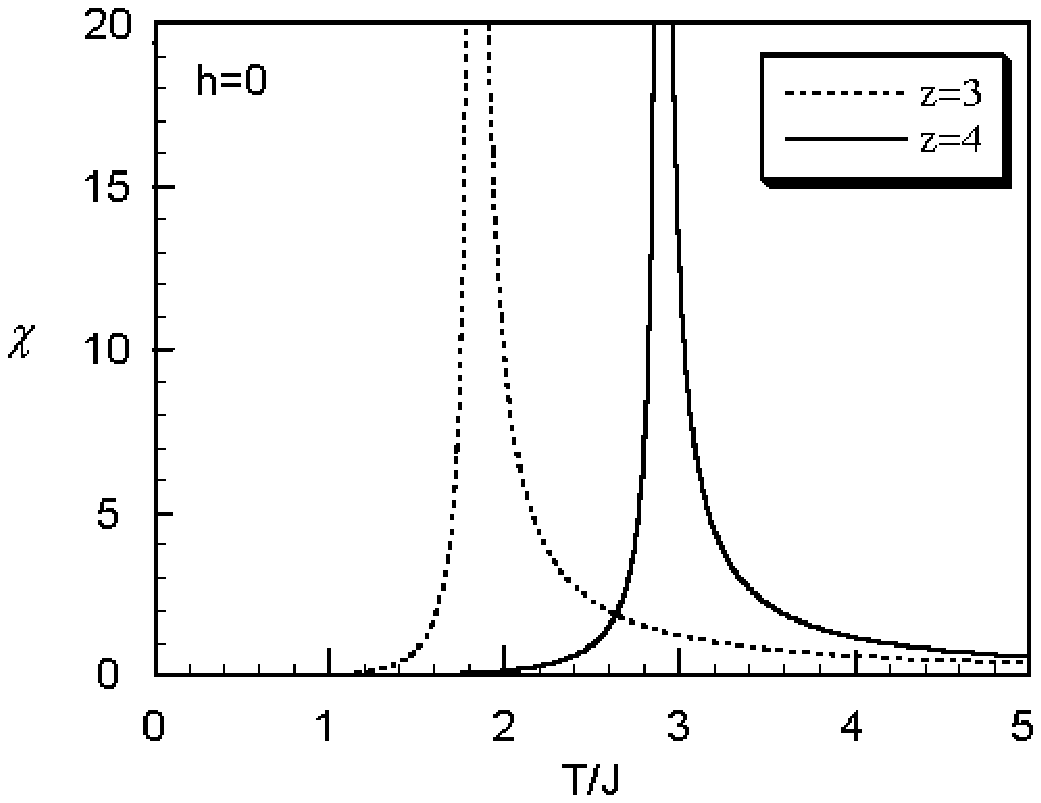}
\caption{The susceptibility $\protect\chi $ is plotted against $T/J$ for $%
z=3 $ and $z=4$ and zero magnetic field.} \label{figura3}
\end{figure}

\begin{figure}[tbph]
\centering\includegraphics*[width=0.5\linewidth]{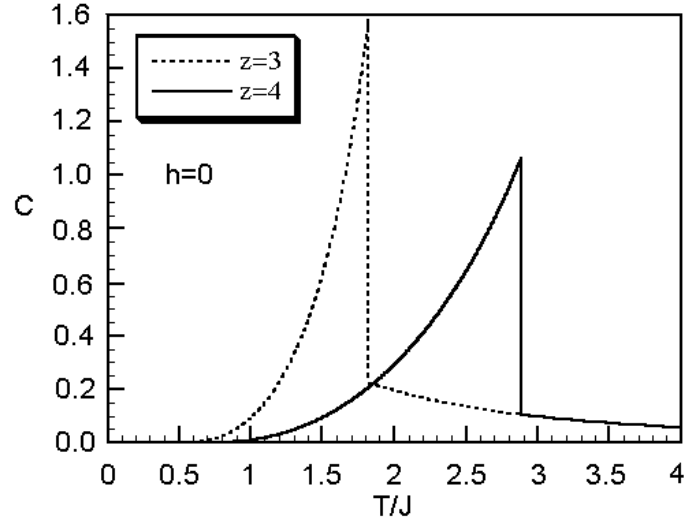}
\caption{The specific heat $C$ is plotted against $T/J$ for $z=3$
and $z=4$ and zero magnetic field.} \label{figura4}
\end{figure}

The temperature dependence of the internal energy $E$, of the free
energy $F$ and of the entropy $S$ is shown in Figs. 5, 6 and 7,
respectively. We
observe the different behaviour at $T_{c}$: $F$ is a smooth function, while $%
E$ and $S$ exhibit a drastic change. This behaviour shows that at
$T_{c}$ we have a second-order phase transition. Also, we note
that above $T_{c}$ the internal energy does not depend on $z$,
while the free energy and the entropy depend on $z$.

\begin{figure}[tbph]
\centering\includegraphics*[width=0.5\linewidth]{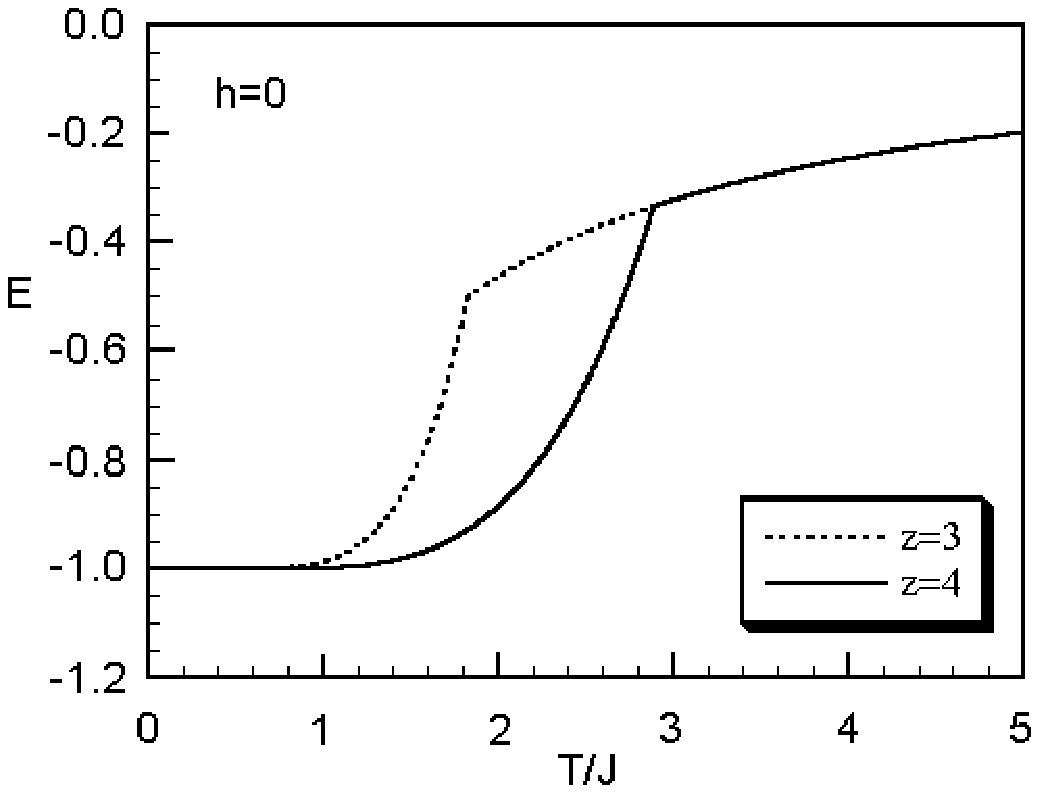}
\caption{The internal energy $E$ is plotted against $T/J$ for
$z=3$ and $z=4$ and zero magnetic field.} \label{figura5}
\end{figure}

\begin{figure}[tbph]
\centering\includegraphics*[width=0.5\linewidth]{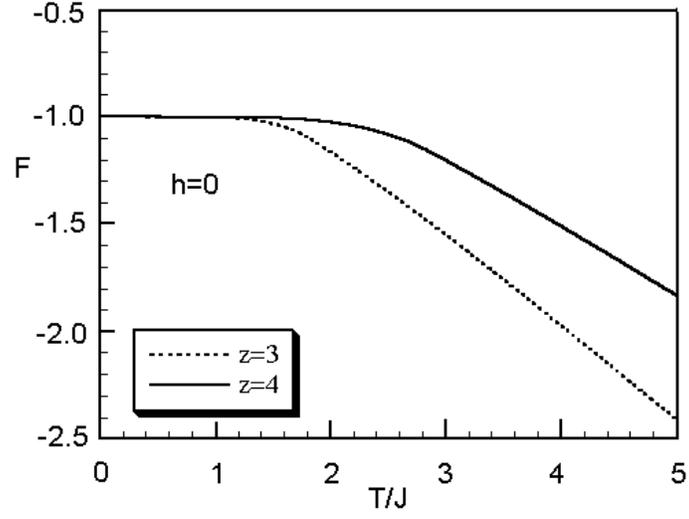}
\caption{The free energy $F$ is plotted against $T/J$ for $z=3$
and $z=4$ and zero magnetic field.} \label{figura6}
\end{figure}

\begin{figure}[tbph]
\centering\includegraphics*[width=0.5\linewidth]{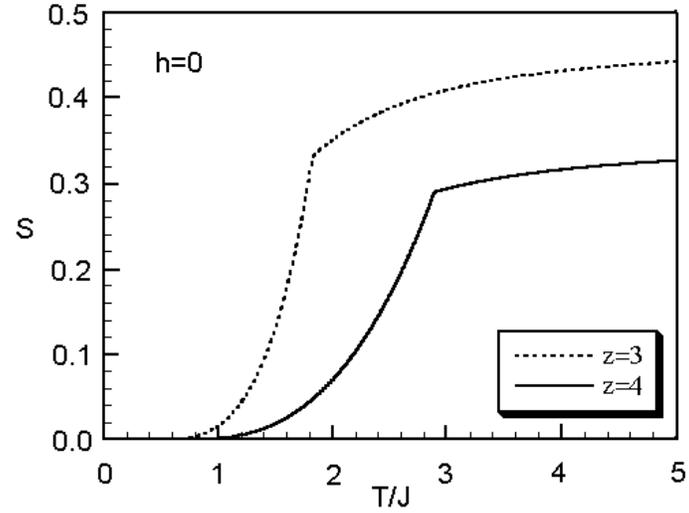}
\caption{The entropy $S$ is plotted against $T/J$ for $z=3$ and
$z=4$ and zero magnetic field.} \label{figura7}
\end{figure}

\begin{figure}[tbph]
\centering\includegraphics*[width=0.5\linewidth]{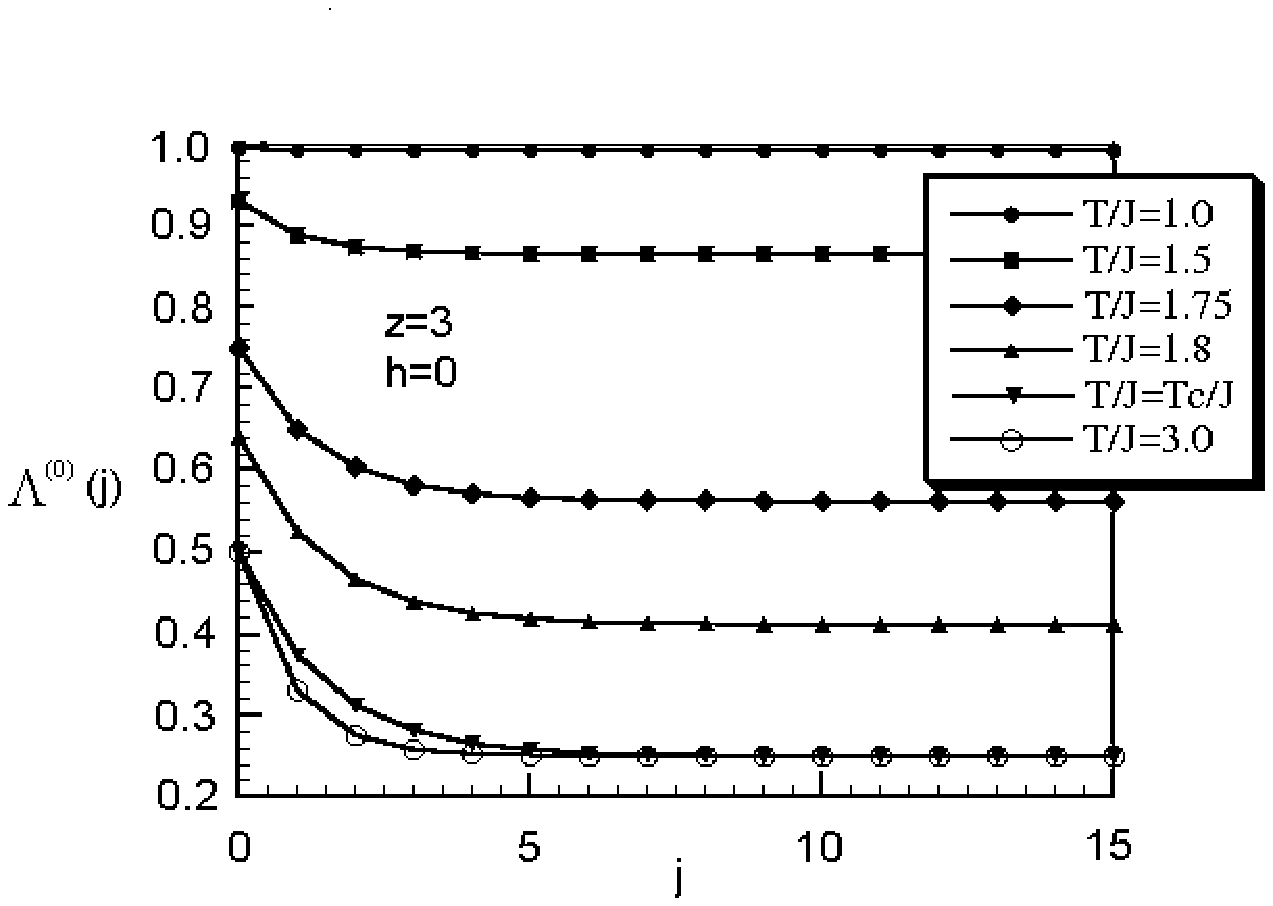}
\caption{The spin correlation function $\Lambda ^{\left( 0\right)
}\left( j\right) $ is plotted against $j$ for $z=3$ and zero
magnetic field.} \label{figura8}
\end{figure}

In Fig. 8 we plot the spin correlation function $\Lambda ^{\left(
0\right) }\left( j\right) =\left\langle n\left( 0\right) n\left(
j\right) \right\rangle $ versus the distance $j$ for $z=3$ and
several values of the temperature, chosen below and above the
critical temperature. We clearly see that a long-range
ferromagnetic order is established below $T_{c}$.

\subsection{Finite magnetic field}

For finite magnetic field the equation (\ref{72}) for the
parameter $X$ can be written as
\begin{equation}
\left( 1-X\right) H\left[ 1+\left( K^{2}-1\right) X\right]
^{z-1}-XK^{z}=0 \label{fmf1}
\end{equation}
where we put $K=e^{2\beta J}$ and $H=e^{2\beta h}$. For $H\neq 1$
this equation does not admit a general solution for any value of
the coordination number, and we must discuss case by case. For
$z=2$ the solution is
\begin{equation}
X=\frac{K^{2}\left( H-1\right)
-2H+K\sqrt{K^{2}+H^{2}K^{2}-2H\left( K^{2}-2\right) }}{2H\left(
K^{2}-1\right) }  \label{fmf2}
\end{equation}
which describes the well known solution of the one-dimensional spin-$\frac{1%
}{2}$ Ising model. The other root of Eq. (\ref{fmf1}) corresponds
to a physically unstable system and is disregarded.

For $z=3$ it is possible to show that there is a critical temperature $%
T_{c}\left( h\right) $, depending on the magnetic field, such that for $%
T<T_{c}\left( h\right) $ there are three real and unequal roots, while for $%
T>T_{c}\left( h\right) $ there is one real root and two conjugate
imaginary roots. $T_{c}\left( h\right) $ is determined by the
following equation:
\begin{equation}
8K^{3}H=K^{4}+18K^{2}-27-\left( K^{2}-9\right)
^{3/2}\sqrt{K^{2}-1}. \label{fmf3}
\end{equation}
For $z=4$ there are four real and unequal roots for $T<T_{c}\left( h\right) $%
, while for $T>T_{c}\left( h\right) $ there are two real unequal
roots and two conjugate imaginary roots.

In Fig. 9 we report $T_{c}\left( h\right) $ as a function of the
magnetic field for $z=3,4$. By increasing $\left| h\right| $,
$T_{c}\left( h\right) $
decreases from the value $k_{B}T_{c}=\frac{2J}{\ln \left( \frac{z}{z-2}%
\right) }$ [cfr. Eq. (\ref{131})] and vanishes at $\left| h\right|
=J\left( z-2\right) $.

\begin{figure}[tbph]
\centering\includegraphics*[width=0.5\linewidth]{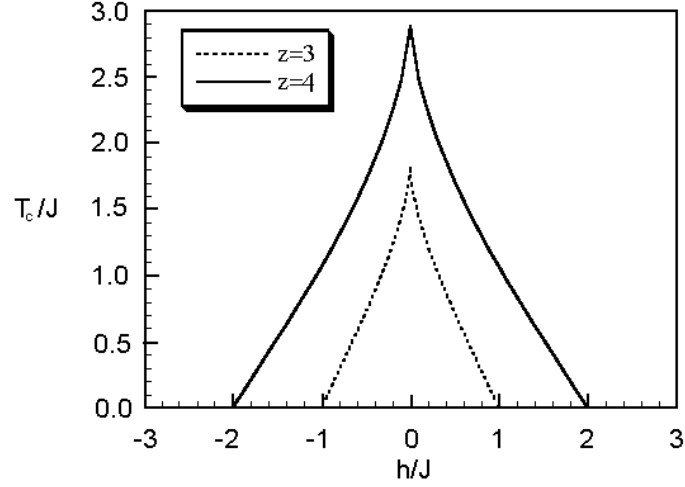}
\caption{The temperature $T_{c}\left( h\right) $ is plotted
against the magnetic field $h$ for $z=3,4$.} \label{figura9}
\end{figure}

For $z=3$, below $T_{c}\left( h\right) $ the three real solutions
have the following behavior: one solution satisfies the
particle-hole symmetry (i. e. $\left\langle n\right\rangle
=\frac{1}{2}$ at $\mu =-2zJ$) but corresponds to an unstable
system (for example the compressibility is negative). The other
two solutions violate the particle-hole symmetry and describe a
finite magnetization for any value of the magnetic field, in a
direction parallel and anti-parallel to $h$, respectively. By
disregarding the unstable solution and by picking up the one which
describes a magnetization in the
direction of the magnetic field, we can assert that in the entire plane $%
\left( h,T\right) $ Eq. (\ref{fmf1}) admits only one physical
solution given by
\[
\begin{array}{cc}
X=2\sqrt{\frac{a}{3}}\cos \left( \theta \right) -\frac{c}{3} &
T<T_{c}\left( h\right)
\end{array}
\]
\begin{equation}
\begin{array}{cc}
X=\sqrt[3]{\frac{b}{2}+\sqrt{\frac{b^{2}}{4}-\frac{a^{3}}{27}}}+\sqrt[3]{%
\frac{b}{2}-\sqrt{\frac{b^{2}}{4}-\frac{a^{3}}{27}}}-\frac{c}{3} &
T>T_{c}\left( h\right)
\end{array}
\label{fmf4}
\end{equation}
where
\begin{equation}
\begin{array}{cc}
a=\frac{K^{3}\left( KH-3\right) }{3H\left( K^{2}-1\right) ^{2}} & b=\frac{%
K^{3}\left( 2HK^{3}-9K^{2}+27\right) }{27H\left( K^{2}-1\right) ^{3}} \\
c=\frac{3-K^{2}}{K^{2}-1} & \theta =\frac{1}{3}\cos ^{-1}\left( \frac{%
3^{3/2}b}{2a^{3/2}}\right)
\end{array}
.  \label{fmf5}
\end{equation}
Similar situation holds for $z=4$. Below $T_{c}\left( h\right) $,
two solutions satisfy the particle-hole symmetry but correspond to
an unstable system; the other two solutions violate the
particle-hole symmetry and describe a finite magnetization,
parallel and anti-parallel to $h$, respectively. Above
$T_{c}\left( h\right) $, among the two solutions, only one has a
physical meaning.

Once $X$ is known, we can calculate the various properties by
using the formulas given in Sections 5 and 6. The behaviour of the
magnetization, the susceptibility and the specific heat as
functions of the temperature is reported in Figs. 10, 12 and 13,
respectively, for several values of the magnetic field. At low
temperatures the system is fully polarized by any finite magnetic
field. By increasing temperature, the magnetization decreases and
tends to zero in the limit $T\rightarrow \infty $. The different
behavior of the magnetization, below and above the critical
temperature $T_{c}$, is shown in Fig. 11, where $m$ is plotted as
a function of the magnetic field for several values of $T$. For
finite $h$ the susceptibility and the specific heat do not exhibit
a discontinuity, there is a peak at a certain temperature $T^{\ast
}$ which increases with $h$.

\begin{figure}[tbph]
\centering\includegraphics*[width=0.5\linewidth]{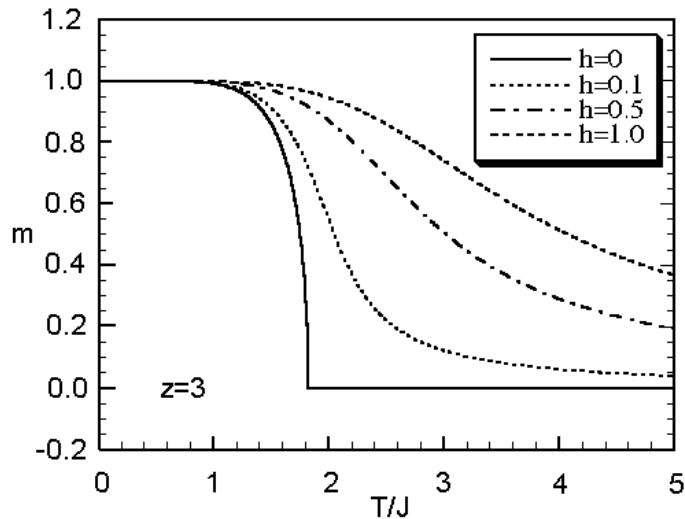}
\caption{The magnetization $m$ is plotted against $T/J$ for $z=3$
and several values of the magnetic field.} \label{figura10}
\end{figure}

\begin{figure}[tbph]
\centering\includegraphics*[width=0.5\linewidth]{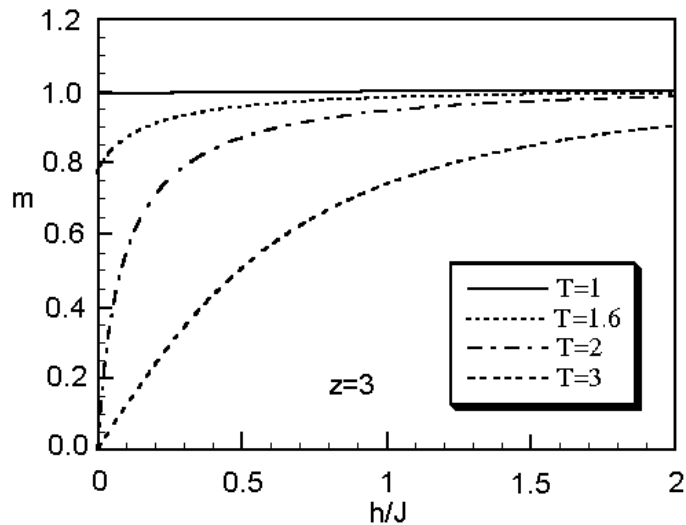}
\caption{The magnetization $m$ is plotted against $h/J$ for $z=3$
and several values of the temperature.} \label{figura11}
\end{figure}

\begin{figure}[tbph]
\centering\includegraphics*[width=0.5\linewidth]{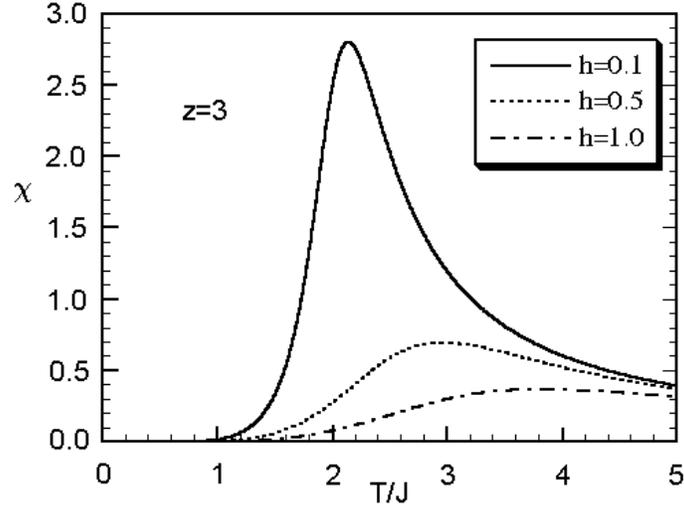}
\caption{The susceptibility $\protect\chi $ is plotted against $T/J$ for $%
z=3 $ and several values of the magnetic field.} \label{figura12}
\end{figure}

\begin{figure}[tbph]
\centering\includegraphics*[width=0.5\linewidth]{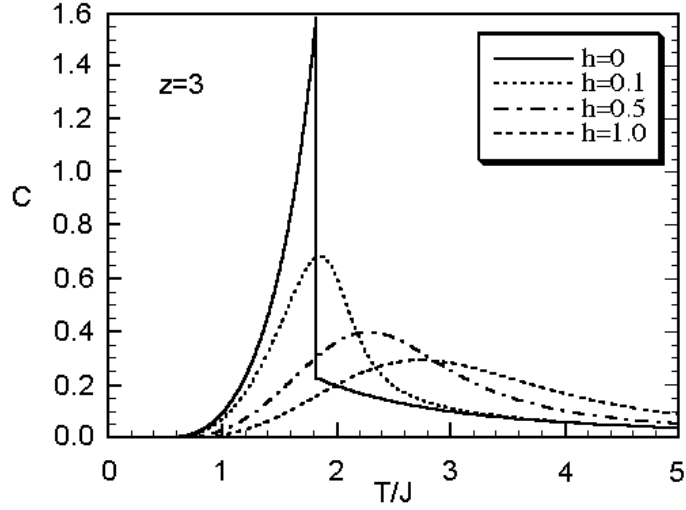}
\caption{The specific heat $C$ is plotted against $T/J$ for $z=3$
and several values of the magnetic field.} \label{figura13}
\end{figure}

\section{Conclusions}

In this paper we have studied the ferromagnetic spin-$\frac{1}{2}$
Ising model in the presence of an external magnetic field on the
Bethe lattice by means of the equations of motion method and of
the Green's function formalism. First, it has been shown that
there exists a mapping between such a model and a fermionic one
built of one species of particles localized on the sites of a
Bethe lattice and interacting via an intersite Coulomb
interaction. Then, an exact solution has been found for the Bethe
lattice with any coordination number $z$ in terms of a finite and
complete set of eigenoperators and eigenvalues of the fermionic
Hamiltonian. This solution allowed us to write exact expressions
for the corresponding Green's function and correlation functions
which depend on a finite set of parameters to be determined in a
self-consistent way. Such parameters have been exactly fixed by
means of algebra constraints. So local and non local correlation
functions have been calculated up to four point ones, together
with the corresponding physical quantities, i.e. the particle
density, the magnetization per site, the susceptibility, the
correlation length, the internal energy per site, the specific
heat and the entropy. All the results are in perfect agreement
with the ones existing in literature. The results are discussed in
great detail for the first two values $z=3,4$ of the coordination
number of the Bethe lattice with and without external magnetic
field. Our procedure allows also to generalize the known results,
as it has been explicitly shown in the case of non local
correlation functions, where new expressions, no previously
reported, for the three-point spin-spin correlation function have
been obtained, together with a general calculation scheme which
can give rise to higher order correlators.

\begin{acknowledgments}
One of the authors (F. M.) wishes to thank Professors G. Baskaran
and N. M. Plakida for stimulating correspondence on the problem of
the Ising model. The authors would like also to thank dr. A.
Avella for stimulating discussions and for a careful reading of
the manuscript.
\end{acknowledgments}

\appendix

\section{Calculation of $\left[ n^{\protect\alpha }\right] ^{k}$%
}

\label{app: primo}

Let us recall the following definition
\begin{equation}
n^{\alpha }\left( i\right) =\frac{1}{z}\left(
n_{1}+n_{2}+...+n_{z}\right) \label{b1}
\end{equation}
where $n_{p}$ ($p=1,...,z$) are the first nearest neighbors of the
site $i$. Then we have to calculate the power
\begin{equation}
\left[ n^{\alpha }\left( i\right) \right]
^{k}=\frac{1}{z^{k}}\left( n_{1}+n_{2}+...+n_{z}\right) ^{k}.
\label{b2}
\end{equation}
By considering the algebraic property $\left[ n_{p}\right] ^{m}=n_{p}$ ($%
m=1,2,...$), after some easy but lengthy calculations, it can be
shown that
\begin{equation}
\left[ n^{\alpha }\left( i\right) \right] ^{k}=\frac{1}{z^{k}}%
\sum_{p=1}^{z}b_{p}^{\left( k\right) }N_{p}^{\left( z\right) }
\label{b3}
\end{equation}
where $N_{p}^{\left( z\right) }$ are the operators
\begin{equation}
N_{p}^{\left( z\right)
}=\sum_{l_{1}<l_{2}<...<l_{p}=1}^{z}n_{l_{1}}n_{l_{2}}...n_{l_{p}}
\label{b5}
\end{equation}
and $b_{p}^{\left( k\right) }$ are some positive integer numbers,
defined as
\begin{equation}
\begin{array}{cc}
b_{1}^{\left( k\right) }=1 &  \\
b_{p}^{\left( k\right)
}=\sum_{l_{1}=p-1}^{k-1}\sum_{l_{2}=p-2}^{l_{1}-1}%
\sum_{l_{3}=p-3}^{l_{2}-1}...\sum_{l_{p-2}=2}^{l_{p-3}-1}%
\sum_{l_{p-1}=1}^{l_{p-2}-1}\left(
\begin{array}{c}
k \\
l_{1}
\end{array}
\right) \left(
\begin{array}{c}
l_{1} \\
l_{2}
\end{array}
\right) \left(
\begin{array}{c}
l_{2} \\
l_{3}
\end{array}
\right) ...\left(
\begin{array}{c}
l_{p-3} \\
l_{p-2}
\end{array}
\right) \left(
\begin{array}{c}
l_{p-2} \\
l_{p-1}
\end{array}
\right) & \left( p>1\right)
\end{array}
.  \label{b6}
\end{equation}
It is important to notice that $b_{p}^{\left( k\right) }=0$ for
$p>k$. In particular for the first values of $p$ we get:
\begin{equation}
\begin{array}{c}
b_{2}^{\left( n\right) }=2\left( 2^{n-1}-1\right) \\
b_{3}^{\left( n\right) }=3\left( 3^{n-1}-2^{n}+1\right) \\
b_{4}^{\left( n\right) }=4\left( 4^{n-1}-3^{n}+3\cdot 2^{n-1}-1\right) \\
b_{5}^{\left( n\right) }=5\left( 5^{n-1}-4^{n}+2\cdot
3^{n}-2^{n+1}+1\right)
\end{array}
.  \label{b7}
\end{equation}
The above results are valid for any lattice with coordination
number $z$.

\section{ Calculation of the coefficients $A_{m}^{\left(
k\right) }$}

Given the results in Appendix A, we can write
\begin{equation}
\left[ n^{\alpha }\left( i\right) \right]
^{k}=\sum_{m=1}^{z}A_{m}^{\left( k\right) }\left[ n^{\alpha
}\left( i\right) \right] ^{m}  \label{a1}
\end{equation}
where the coefficients $A_{m}^{\left( k\right) }$ are some
rational numbers which must satisfy the relations
\begin{equation}
\begin{array}{c}
\sum_{m=1}^{z}A_{m}^{\left( k\right) }=1 \\
\begin{array}{cc}
A_{m}^{\left( k\right) }=\delta _{m,k} & \left( k=1,...,z\right)
\end{array}
\end{array}
.  \label{a2}
\end{equation}
The first relation follows by putting $n^{\alpha }\left( i\right)
=1$ while the second can be derived by considering the case $1\leq
k\leq z$. Indeed we must calculate the coefficients $A_{m}^{\left(
k\right) }$ only for $k\geq z+1$ and $m=1,...,z$. By noting that
for $k\geq z+2$ we can write
\begin{equation}
\sum_{m=1}^{z}A_{m}^{\left( k\right) }\left[ n^{\alpha }\left(
i\right) \right] ^{m}=\sum_{m=1}^{z}A_{m}^{\left( k-1\right)
}\left[ n^{\alpha }\left( i\right) \right] ^{m+1},  \label{a4}
\end{equation}
the following recursion rule can be established
\begin{equation}
\begin{array}{ccc}
A_{m}^{\left( k\right) }=A_{m-1}^{\left( k-1\right)
}+A_{z}^{\left( k-1\right) }A_{m}^{\left( z+1\right) } & \left(
m=1,...,z\right) & A_{0}^{\left( k-1\right) }=0
\end{array}
.  \label{a6}
\end{equation}
This rule implies that we must calculate only the $z$ coefficients $%
A_{m}^{\left( z+1\right) }$, ($m=1,...,z$), which can be done by
means of the relation (\ref{a1}) evaluated for $k=z+1$:
\begin{equation}
\left[ n^{\alpha }\left( i\right) \right]
^{z+1}=\sum_{m=1}^{z}A_{m}^{\left( z+1\right) }\left[ n^{\alpha
}\left( i\right) \right] ^{m}.  \label{a7}
\end{equation}
By using the results given in Appendix A to rewrite (\ref{a7}), we
obtain the equation
\begin{equation}
\sum_{k=1}^{z}b_{k}^{\left( z+1\right) }N_{k}^{\left( z\right)
}=\sum_{m=1}^{z}A_{m}^{\left( z+1\right)
}z^{z+1-m}\sum_{k=1}^{m}b_{k}^{\left( m\right) }N_{k}^{\left(
z\right) }, \label{a9}
\end{equation}
which, by noting that the operators $N_{k}^{\left( z\right) }$ are
linearly independent, takes the form
\begin{equation}
\begin{array}{cc}
\sum_{m=k}^{z}A_{m}^{\left( z+1\right) }z^{z+1-m}b_{k}^{\left(
m\right) }-b_{k}^{\left( z+1\right) }=0 & \left( k=1,...,z\right)
\end{array}
.  \label{a11}
\end{equation}
Such equations give rise to the iterative solution
\begin{equation}
\begin{array}{c}
A_{z}^{\left( z+1\right) }=\frac{b_{z}^{\left( z+1\right)
}}{zb_{z}^{\left(
z\right) }} \\
A_{z-1}^{\left( z+1\right) }=\frac{1}{z^{2}b_{z-1}^{\left( z-1\right) }}%
\left[ b_{z-1}^{\left( z+1\right) }-A_{z}^{\left( z+1\right)
}zb_{z-1}^{\left( z\right) }\right] \\
A_{z-2}^{\left( z+1\right) }=\frac{1}{z^{3}b_{z-2}^{\left( z-2\right) }}%
\left[ b_{z-2}^{\left( z+1\right) }-A_{z-1}^{\left( z+1\right)
}z^{2}b_{z-2}^{\left( z-1\right) }-A_{z}^{\left( z+1\right)
}zb_{z-2}^{\left( z\right) }\right] \\
\vdots \\
A_{1}^{\left( z+1\right) }=\frac{1}{z^{z}}\left[ 1-A_{2}^{\left(
z+1\right) }z^{z-1}-A_{3}^{\left( z+1\right)
}z^{z-2}-...-A_{z-2}^{\left( z+1\right) }z^{3}-A_{z-1}^{\left(
z+1\right) }z^{2}-A_{z}^{\left( z+1\right) }z\right]
\end{array}
.  \label{a12}
\end{equation}
As an example, we give the values of the coefficients
$A_{m}^{\left( k\right) }$ for the first values of $z$
\begin{equation}
\begin{array}{ccccccc}
z=2: & A_{1}^{\left( 3\right) }=-\frac{1}{2} & A_{2}^{\left( 3\right) }=%
\frac{3}{2} &  &  &  &  \\
z=3: & A_{1}^{\left( 4\right) }=\frac{2}{9} & A_{2}^{\left( 4\right) }=-%
\frac{11}{9} & A_{3}^{\left( 4\right) }=2 &  &  &  \\
z=4: & A_{1}^{\left( 5\right) }=-\frac{3}{32} & A_{2}^{\left( 5\right) }=%
\frac{25}{32} & A_{3}^{\left( 5\right) }=-\frac{35}{16} &
A_{4}^{\left(
5\right) }=\frac{5}{2} &  &  \\
z=5: & A_{1}^{\left( 6\right) }=\frac{24}{625} & A_{2}^{\left( 6\right) }=-%
\frac{274}{625} & A_{3}^{\left( 6\right) }=\frac{9}{5} &
A_{4}^{\left(
6\right) }=-\frac{17}{5} & A_{5}^{\left( 6\right) }=3 &  \\
z=6: & A_{1}^{\left( 7\right) }=-\frac{5}{324} & A_{2}^{\left( 7\right) }=%
\frac{49}{216} & A_{3}^{\left( 7\right) }=-\frac{203}{162} &
A_{4}^{\left( 7\right) }=\frac{245}{72} & A_{5}^{\left( 7\right)
}=-\frac{175}{36} & A_{6}^{\left( 7\right) }=\frac{7}{2}
\end{array}
.  \label{a13}
\end{equation}

\section{Calculation of $\left\langle \left[ n^{\protect\alpha
}\left( 0\right) \right] ^{k}\right\rangle _{0}$}

By means of the results given in Appendix A we can write that:
\begin{equation}
\left\langle \left[ n^{\alpha }\left( 0\right) \right]
^{k}\right\rangle _{0}=\frac{1}{z^{k}}\sum_{p=1}^{z}b_{p}^{\left(
k\right) }\left\langle N_{p}^{\left( z\right) }\right\rangle _{0}.
\label{c1}
\end{equation}
The particular topology of the Bethe lattice allows us to decouple
the correlation functions in the $H_{0}$-representation (see Eq.
(\ref{52})) and to obtain
\begin{equation}
\left\langle N_{p}^{\left( z\right) }\right\rangle
_{0}=\sum_{l_{1}<l_{2}<...<l_{p}=1}^{z}\left\langle
n_{l_{1}}n_{l_{2}}...n_{l_{p}}\right\rangle _{0}=\left[
\left\langle
n^{\alpha }\left( 0\right) \right\rangle _{0}\right] ^{p}%
\sum_{l_{1}<l_{2}<...<l_{p}=1}^{z}=\left(
\begin{array}{c}
z \\
p
\end{array}
\right) \left[ \left\langle n^{\alpha }\left( 0\right) \right\rangle _{0}%
\right] ^{p}.  \label{c2}
\end{equation}
It follows that
\begin{equation}
\left\langle \left[ n^{\alpha }\left( 0\right) \right]
^{k}\right\rangle _{0}=\frac{1}{z^{k}}\sum_{p=1}^{z}\left(
\begin{array}{c}
z \\
p
\end{array}
\right) b_{p}^{\left( k\right) }X^{p}  \label{c3}
\end{equation}
where $X=\left\langle n^{\alpha }\left( 0\right) \right\rangle
_{0}$ (see Eq. (\ref{55})).

\section{Calculation of $\protect\lambda ^{\left( k\right) }$
and $\protect\kappa ^{\left( k\right) }$}

In this Appendix we will calculate the local correlators $\lambda
^{\left( k\right) }$ and $\kappa ^{\left( k\right) }$ in terms of
the parameter $X$ defined in Eq. (\ref{55}). We recall the
definitions given in Section 5:
\begin{equation}
\begin{array}{c}
\lambda ^{\left( k\right) }=\left\langle n\left( 0\right) \left[
n^{\alpha
}\left( 0\right) \right] ^{k}\right\rangle  \\
\kappa ^{\left( k\right) }=\left\langle \left[ n^{\alpha }\left(
0\right) \right] ^{k}\right\rangle
\end{array}
.  \label{d1}
\end{equation}
Let us start with $\lambda ^{\left( k\right) }$ which, according to (\ref{43}%
), can be written as
\begin{equation}
\lambda ^{\left( k\right) }=\frac{\left\langle n\left( 0\right)
\left[
n^{\alpha }\left( 0\right) \right] ^{k}e^{-\beta H_{I}}\right\rangle _{0}}{%
\left\langle e^{-\beta H_{I}}\right\rangle _{0}}.  \label{d2}
\end{equation}
By recalling Eq. (\ref{79}) and the recursion rule (\ref{b3}), we
have
\begin{equation}
\left\langle n\left( 0\right) \left[ n^{\alpha }\left( 0\right)
\right]
^{k}e^{-\beta H_{I}}\right\rangle _{0}=\frac{1}{z^{k}}\sum_{p=1}^{z}b_{p}^{%
\left( k\right) }\left\langle n\left( 0\right) \right\rangle
_{0}\left\langle N_{p}^{\left( z\right) }\prod_{i=1}^{z}\left[ 1+An_{i}%
\right] \right\rangle _{0}.  \label{d4}
\end{equation}
Furthermore, by recalling Eq. (\ref{b5}) it is immediate to show
that
\begin{equation}
\left\langle N_{p}^{\left( z\right) }\prod_{i=1}^{z}\left[
1+An_{i}\right] \right\rangle _{0}=\left(
\begin{array}{c}
z \\
p
\end{array}
\right) \left( 1+A\right) ^{p}X^{p}\left( 1+AX\right) ^{z-p}.
\label{d5}
\end{equation}
By putting (\ref{d4}) and (\ref{d5}) into (\ref{d2}) and recalling
the results (\ref{78}) and (\ref{80}) we obtain for $k\geq 1$
\begin{equation}
\lambda ^{\left( k\right) }=\frac{\left( 1-X\right) }{1+AX^{2}}e^{\beta \mu }%
\frac{1}{z^{k}}\sum_{p=1}^{z}\left(
\begin{array}{c}
z \\
p
\end{array}
\right) b_{p}^{\left( k\right) }\left( 1+A\right) ^{p}X^{p}\left(
1+AX\right) ^{z-p}.  \label{d6}
\end{equation}
Use of the equation (\ref{72}) for the parameter $X$ allows us to rewrite (%
\ref{d6}) under the form
\begin{equation}
\lambda ^{\left( k\right)
}=\frac{1}{1+AX^{2}}\sum_{p=1}^{k}a_{p}^{\left( z,k\right)
}\frac{X^{p+1}\left( 1+A\right) ^{p}}{\left( 1+AX\right) ^{p-1}}
\label{d8}
\end{equation}
where
\begin{equation}
a_{p}^{\left( z,k\right) }=\frac{1}{z^{k}}b_{p}^{\left( k\right)
}\left(
\begin{array}{c}
z \\
p
\end{array}
\right) .  \label{d9}
\end{equation}
In order to calculate $\kappa ^{\left( k\right) }$, let us start
from the equation:
\begin{equation}
\kappa ^{\left( k\right) }=C_{1,k+1}+\lambda ^{\left( k\right) }.
\label{d10}
\end{equation}
By putting together (\ref{84}) and (\ref{d8}), we obtain:
\begin{equation}
\kappa ^{\left( k\right)
}=\frac{1}{1+AX^{2}}\sum_{p=1}^{z}a_{p}^{\left(
z,k\right) }X^{p}\left[ \left( 1-X\right) +\frac{X\left( 1+A\right) ^{p}}{%
\left( 1+AX\right) ^{p-1}}\right] .  \label{d11}
\end{equation}
In particular, from (\ref{d8}) and (\ref{d11}) we have
\begin{eqnarray}
\lambda ^{\left( 1\right) } &=&\frac{X^{2}\left( 1+A\right)
}{1+AX^{2}}
\nonumber \\
\lambda ^{\left( 2\right) } &=&\frac{1}{z}\lambda ^{\left( 1\right) }+\frac{%
z-1}{z}\frac{X^{3}\left( 1+A\right) ^{2}}{\left( 1+AX^{2}\right)
\left( 1+AX\right) }  \label{d12}
\end{eqnarray}
and
\begin{eqnarray}
\kappa ^{\left( 1\right) } &=&\frac{X\left( 1+AX\right)
}{1+AX^{2}}
\nonumber \\
\kappa ^{\left( 2\right) } &=&\frac{1}{z}\kappa ^{\left( 1\right) }+\frac{z-1%
}{z}\frac{X^{2}}{1+AX^{2}}\left[ \left( 1-X\right) +\frac{X\left(
1+A\right) ^{2}}{\left( 1+AX\right) }\right] .  \label{d13}
\end{eqnarray}
In closing this Appendix, we note the following useful relations
\begin{eqnarray}
\kappa ^{\left( 2\right) } &=&\frac{1}{z}n+\frac{z-1}{z}\left[ n^{2}+\frac{%
\left( n^{2}-\lambda ^{\left( 1\right) }\right) ^{2}}{n\left( 1-n\right) }%
\right]   \nonumber \\
\kappa ^{\left( 2\right) }-\lambda ^{\left( 2\right) }
&=&\frac{1}{z}\left( n-\lambda ^{\left( 1\right) }\right)
+\frac{z-1}{z}\frac{\left( n-\lambda
^{\left( 1\right) }\right) ^{2}}{\left( 1-n\right) }  \label{d14} \\
\lambda ^{\left( 2\right) } &=&\frac{1}{z}\lambda ^{\left( 1\right) }+\frac{%
z-1}{z}\frac{\left[ \lambda ^{\left( 1\right) }\right] ^{2}}{n}.
\nonumber
\end{eqnarray}
where we used $n=\kappa ^{\left( 1\right) }$.

\section{Calculation of $\Lambda ^{\left( 0\right) }\left(
j\right) $ and $K^{\left( 1\right) }\left( j\right) $}

Let us recall the following definitions given in Section 6:
\begin{equation}
\begin{array}{c}
K^{\left( k\right) }\left( j\right) =\left\langle \left[ n^{\alpha
}\left(
0\right) \right] ^{k}n\left( j\right) \right\rangle  \\
\Lambda ^{\left( k\right) }\left( j\right) =\left\langle n\left(
0\right) \left[ n^{\alpha }\left( 0\right) \right] ^{k}n\left(
j\right) \right\rangle
\end{array}
\label{e1}
\end{equation}
where $j$ is a site at a distance of $j$ ($j\geq 2$) steps from
the central site. Let us make for simplicity the choice that $j$
belongs to the $z$-th subtree (but any subtree can be chosen) and
let us define the correlation function of the composite fields as
\begin{equation}
\begin{array}{cc}
D^{\left( k\right) }\left( j\right) =\left\langle c\left( 0\right)
c^{\dagger }\left( 0\right) \left[ n^{\alpha }\left( 0\right)
\right] ^{k-1}n\left( j\right) \right\rangle  & \left( k\geq
1\right)
\end{array}
.  \label{e2}
\end{equation}
By means of the commutation relations we note that
\begin{equation}
D^{\left( k\right) }\left( j\right) =K^{\left( k-1\right) }\left(
j\right) -\Lambda ^{\left( k-1\right) }\left( j\right) ,
\label{e3}
\end{equation}
while, by using (\ref{43}) and the algebraic relation (\ref{50}),
we have
\begin{equation}
D^{\left( k\right) }\left( j\right) =\frac{\left\langle c\left(
0\right) c^{\dagger }\left( 0\right) \left[ n^{\alpha }\left(
0\right) \right] ^{k-1}n\left( j\right) \right\rangle
_{0}}{\left\langle e^{-\beta H_{I}}\right\rangle _{0}}.
\label{e4}
\end{equation}
By means of the properties of the correlation functions in the $H_{0}$%
-representation (see Eq. (\ref{52})) we get:
\begin{eqnarray}
D^{\left( k\right) }\left( j\right)  &=&\frac{\left\langle c\left(
0\right) c^{\dagger }\left( 0\right) \right\rangle
_{0}}{\left\langle e^{-\beta H_{I}}\right\rangle _{0}}\left\langle
\left[ n^{\alpha }\left( 0\right) \right] ^{k-1}\left[ n\left(
j\right) \right] \right\rangle _{0}  \nonumber
\\
&=&C_{1,1}\left\langle \left[ n^{\alpha }\left( 0\right) \right]
^{k-1}\left[ n\left( j\right) \right] \right\rangle _{0}=\left(
1-n\right) \left\langle \left[ n^{\alpha }\left( 0\right) \right]
^{k-1}\left[ n\left( j\right) \right] \right\rangle _{0},
\label{e5}
\end{eqnarray}
where we used Eq. (\ref{58}). By putting together (\ref{e3}) and
(\ref{e5}) we obtain
\begin{equation}
K^{\left( k-1\right) }\left( j\right) -\Lambda ^{\left( k-1\right)
}\left( j\right) =\left( 1-n\right) \left\langle \left[ n^{\alpha
}\left( 0\right) \right] ^{k-1}\left[ n\left( j\right) \right]
\right\rangle _{0},  \label{e6}
\end{equation}
which in particular for $k=1$ reads
\begin{equation}
n-\Lambda ^{\left( 0\right) }\left( j\right) =\left( 1-n\right)
\left\langle n\left( j\right) \right\rangle _{0}.  \label{e7}
\end{equation}
Let us notice that $\left\langle n\left( j\right) \right\rangle
_{0}$ depends on the site $j$ because the $H_{0}$-representation
lacks of
translational invariance. Let us now start to calculate the function $%
\Lambda ^{\left( 0\right) }\left( j\right) =\left\langle n\left(
0\right) n\left( j\right) \right\rangle $, which in the
$H_{0}$-representation can be written as
\begin{equation}
\Lambda ^{\left( 0\right) }\left( j\right) =\left\langle n\left(
0\right) n\left( j\right) \right\rangle =\frac{\left\langle
n\left( 0\right) n\left( j\right) e^{-\beta H_{I}}\right\rangle
_{0}}{\left\langle e^{-\beta H_{I}}\right\rangle _{0}}.
\label{e8}
\end{equation}
By recalling that (cfr. (\ref{79})) $e^{-\beta
H_{I}}=\prod_{i=1}^{z}\left[ 1+An\left( 0\right) n_{i}\right] $ we
have
\begin{equation}
\left\langle n\left( 0\right) n\left( j\right) e^{-\beta
H_{I}}\right\rangle _{0}=\left\langle n\left( 0\right)
\right\rangle _{0}\left( 1+AX\right) ^{z-1}\left[ \left\langle
n\left( j\right) \right\rangle _{0}+A\left\langle n_{z}n\left(
j\right) \right\rangle _{0}\right] .  \label{e9}
\end{equation}
Putting (\ref{e9}) into (\ref{e8}), recalling the results (\ref{78}) and (%
\ref{80}), using the equation (\ref{72}) for the parameter $X$, we
obtain
\begin{equation}
\Lambda ^{\left( 0\right) }\left( j\right)
=\frac{X}{1+AX^{2}}\left[ \left\langle n\left( j\right)
\right\rangle _{0}+A\left\langle n_{z}n\left( j\right)
\right\rangle _{0}\right] .  \label{e13}
\end{equation}
By combining (\ref{e7}) and (\ref{e13}) we can express the unknown
correlation functions $\left\langle n\left( j\right) \right\rangle
_{0}$ and $\left\langle n_{z}n\left( j\right) \right\rangle _{0}$
in terms of the two-point correlation function $\Lambda ^{\left(
0\right) }\left( j\right) $ as follows
\begin{eqnarray}
\left\langle n\left( j\right) \right\rangle _{0}
&=&\frac{1}{\left( 1-n\right) }\left[ n-\Lambda ^{\left( 0\right)
}\left( j\right) \right]
\nonumber \\
\left\langle n_{z}n\left( j\right) \right\rangle _{0} &=&\left[ \frac{%
1+AX^{2}}{AX}+\frac{1}{A}\frac{1}{\left( 1-n\right) }\right]
\Lambda ^{\left( 0\right) }\left( j\right)
-\frac{1}{A}\frac{n}{\left( 1-n\right) }. \label{e14}
\end{eqnarray}
Let us now calculate the function $\Lambda ^{\left( 1\right)
}\left( j\right) =\left\langle n\left( 0\right) n^{\alpha }\left(
0\right) n\left( j\right) \right\rangle $ by following the same
procedure we adopted for the calculation of $\Lambda ^{\left(
0\right) }\left( j\right) $. Recalling (\ref {78}) and (\ref{80}),
and the basic equation (\ref{72}) we get
\begin{eqnarray}
\Lambda ^{\left( 1\right) }\left( j\right)  &=&\frac{\left( z-1\right) }{z}%
\frac{X^{2}\left( 1+A\right) }{\left( 1+AX\right) \left( 1+AX^{2}\right) }%
\left[ \left\langle n\left( j\right) \right\rangle
_{0}+A\left\langle
n_{z}n\left( j\right) \right\rangle _{0}\right]   \nonumber \\
&&+\frac{1}{z}\frac{X\left( 1+A\right) }{1+AX^{2}}\left\langle
n_{z}n\left( j\right) \right\rangle _{0}.  \label{e18}
\end{eqnarray}
Now, by using the relations in (\ref{e14}) it is possible to
express the correlation function $\Lambda ^{\left( 1\right)
}\left( j\right) $ in terms of $\Lambda ^{\left( 0\right) }\left(
j\right) $:
\begin{equation}
\Lambda ^{\left( 1\right) }\left( j\right)
=-\frac{1}{z}\frac{X\left( 1+A\right)
}{1+AX^{2}}\frac{1}{A}\frac{n}{\left( 1-n\right) }+\frac{\left(
z-1\right) }{z}\frac{X\left( 1+A\right) }{\left( 1+AX\right)
}\Lambda
^{\left( 0\right) }\left( j\right) +\frac{1}{z}\frac{\left( 1+A\right) }{%
A\left( 1-X\right) }\Lambda ^{\left( 0\right) }\left( j\right) .
\label{e19}
\end{equation}
By using the relation (\ref{e6}) for $k=2$ we are now in position
to calculate the function $K^{\left( 1\right) }\left( j\right) $
as
\begin{equation}
K^{\left( 1\right) }\left( j\right) =\Lambda ^{\left( 1\right)
}\left( j\right) +\left( 1-n\right) \left\langle n^{\alpha }\left(
0\right) n\left( j\right) \right\rangle _{0},  \label{e20}
\end{equation}
which, by observing that
\begin{equation}
\left\langle n^{\alpha }\left( 0\right) n\left( j\right) \right\rangle _{0}=%
\frac{1}{z}\left\langle n_{z}n\left( j\right) \right\rangle _{0}+\frac{z-1}{z%
}X\left\langle n\left( j\right) \right\rangle _{0},  \label{e21}
\end{equation}
and recalling (\ref{e14}) and (\ref{e19}), becomes
\begin{equation}
K^{\left( 1\right) }\left( j\right) =\frac{1}{zp}\left[ \Lambda
^{\left( 0\right) }\left( j\right) -n^{2}\right] +\frac{\left(
z-1\right) p}{z}\left[ \Lambda ^{\left( 0\right) }\left( j\right)
-n^{2}\right] +n^{2}  \label{e23}
\end{equation}
where we made use of the relations (\ref{82}) and (\ref{92}).

Recalling now the recursion relation (cfr. (\ref{111}))
\begin{equation}
K^{\left( 1\right) }\left( j\right) =\frac{1}{z}\Lambda ^{\left(
0\right) }\left( j-1\right) +\frac{z-1}{z}\Lambda ^{\left(
0\right) }\left( j+1\right) \label{e25}
\end{equation}
and putting that together with (\ref{e23}), we finally obtain the
relevant recurrence relation
\begin{equation}
G\left( j+1\right) -pG\left( j\right) =\frac{1}{p\left( z-1\right)
}\left[ G\left( j\right) -pG\left( j-1\right) \right]  \label{e26}
\end{equation}
where we defined
\begin{equation}
G\left( j\right) =\Lambda ^{\left( 0\right) }\left( j\right)
-n^{2}=\left\langle n\left( 0\right) n\left( j\right)
\right\rangle -n^{2}. \label{e27}
\end{equation}

\end{document}